\documentclass[11pt]{elsarticle}

\usepackage{lineno,hyperref}

\usepackage{float}
\usepackage{graphicx}
\usepackage{subfig}
\usepackage{mathrsfs}
\usepackage{graphics}
\usepackage{rotating}

\usepackage{float}
\usepackage{soul}

\usepackage{amsmath, commath}
\usepackage{amsfonts}
\usepackage{mathrsfs}
\usepackage{amsmath}
\usepackage[table]{xcolor}

\usepackage{optidef}

\usepackage{optidef}
\usepackage{graphicx}
\usepackage{float}
\usepackage{graphicx}
\usepackage{subfig}
\usepackage{mathrsfs}
\usepackage{float}
\usepackage{amsmath}
\usepackage{amsfonts}
\usepackage{booktabs}
\usepackage{natbib}
\usepackage{bbm}
\usepackage{algorithm}
\usepackage[noend]{algpseudocode}
\usepackage{multirow}
\algdef{SE}[DOWHILE]{Do}{doWhile}{\algorithmicdo}[1]{\algorithmicwhile\ #1}

\usepackage[letterpaper, portrait, margin=1in]{geometry}

\usepackage{booktabs}
\usepackage{tabularx}

\usepackage{xcolor}
\hypersetup{
    colorlinks,
    linkcolor={red!50!red},
    citecolor={blue!50!blue},
    urlcolor={blue!80!blue}
}

\usepackage{optidef}

\usepackage{amsthm}

\theoremstyle{definition}

\makeatletter

\usepackage{newtxtext} % use times new roman font style

\makeatletter
\newcommand*{\centerfloat}{%
  \parindent \z@
  \leftskip \z@ \@plus 1fil \@minus \marginparwidth
  \rightskip \leftskip
  \parfillskip \z@skip}
\makeatother

\usepackage{amssymb,amsmath,caption}
\usepackage[hang,flushmargin]{footmisc}
\newcommand{\algorithmfootnote}[2][\footnotesize]{
  \let\old@algocf@finish\@algocf@finish% Store algorithm finish macro
  \def\@algocf@finish{\old@algocf@finish% Update finish macro to insert "footnote"
    \leavevmode\rlap{\begin{minipage}{\linewidth}
    #1#2
    \end{minipage}}%
  }%
}

\usepackage{setspace}
\usepackage{natbib}
\usepackage{algorithm}
\usepackage[noend]{algpseudocode}
\algdef{SE}[DOWHILE]{Do}{doWhile}{\algorithmicdo}[1]{\algorithmicwhile\ #1}

\usepackage{appendix}
\usepackage{array}
\usepackage{threeparttable}
\usepackage{multirow}
\usepackage{rotating}
\usepackage{makecell}

\journal{}

\biboptions{authoryear}

\begin{document}
\begin{frontmatter}

\title{
        Large Scale Multi-GPU Based Parallel Traffic Simulation for\\
        Accelerated Traffic Assignment and Propagation%
}

% \title{Appendix: Methodology for probabilistic Inference}


\author[label1]{Xuan Jiang\corref{mycorrespondingauthor}}
\author[label1]{Raja Sengupta}
\author[label2]{James Demmel}
\author[label4]{Samuel Williams}
\address[label1]{Department of Civil and Environmental Engineering, University of California, Berkeley, CA 94709}
\address[label2]{Department of Electrical Engineering and Computer Science, University of California, Berkeley, CA 94709}
\address[label4]{Performance and Algorithms Research Group, Lawrence Berkeley National Laboratory, Berkeley, CA 94720}

\cortext[mycorrespondingauthor]{Corresponding author}

\begin{abstract}
Traffic simulation is a critical tool for congestion analysis, travel time estimation, and route optimization in urban planning, benefiting navigation apps, transportation network companies, and state agencies. Traditionally, traffic micro-simulation frameworks are based on road segments and can only support a limited number of main roads. Efficient traffic simulation on a regional scale remains a significant challenge due to the complexity of urban mobility and the large scale of spatiotemporal data. This paper introduces a Large Scale Multi-GPU Parallel Computing based Regional Scale Traffic Simulation Framework (LPSim), which leverages graphical processing unit (GPU) parallel computing to address these challenges. LPSim utilizes a multi-GPU architecture to simulate extensive and dynamic traffic networks with high fidelity and reduced computation time. Using the parallel processing capabilities of GPUs, LPSim can perform tens of millions of individual vehicle dynamics simulations simultaneously, significantly outperforming traditional CPU-based approaches. The framework is designed to be scalable and can easily accommodate the increasing complexity of traffic simulations. We present the theory behind GPU-based traffic simulation, the architecture of single- and multi-GPU based simulations, and the graph partition strategies that enhance computation resource load balance. Our experimental results demonstrate the effectiveness of LPSim in simulating large-scale traffic scenarios. LPSim is capable of completing simulations of 2.82 million trips in just 6.28 minutes on a single GPU machine equipped with 5120 CUDA cores (Tesla V100-SXM2). Furthermore, utilizing a Google Cloud instance with two NVIDIA V100 GPUs, which collectively offer 10240 CUDA cores, LPSim successfully simulates 9.01 million trips within 21.16 minutes. We further tested our simulator with the same demand on dual NVIDIA A100-PCIE-40GB GPUs, which finished the simulation in 0.0398 hours, approximately 113 times faster than the same simulation scenario running on an Intel(R) Xeon(R) Gold 6326 CPU @ 2.90GHz, which takes 4.49 hours to complete. This performance not only demonstrates its speed and scalability advantages over traditional simulation techniques but also highlights LPSim's unique position as the first traffic simulation framework that is scalable for both single- and multiple-GPU configurations. Consequently, LPSim provides an invaluable tool for individuals and extensive research teams alike, enabling the acquisition of large-scale traffic simulation results in a time-efficient manner. LPSim code is available at: \url{https://github.com/Xuan-1998/LPSim}

\end{abstract}


\end{frontmatter}

% \linenumbers % LINE NUMBER

\section{Introduction}
Traffic simulation, a powerful tool that bridges the areas of computer science and traffic engineering, plays a crucial role in understanding how urban mobility works \cite{zomer2015meta}\cite{jiang2023simulating}. By simulating and recreating traffic scenarios in virtual environments, traffic simulation systems offer a robust platform for researchers, engineers, and policymakers to analyze, design, and experiment strategies to avoid the risks and costs of real-world trials \cite{krautter1999traffic}. 

\begin{figure}[!htbp]
\centerline{\includegraphics[width=\columnwidth]{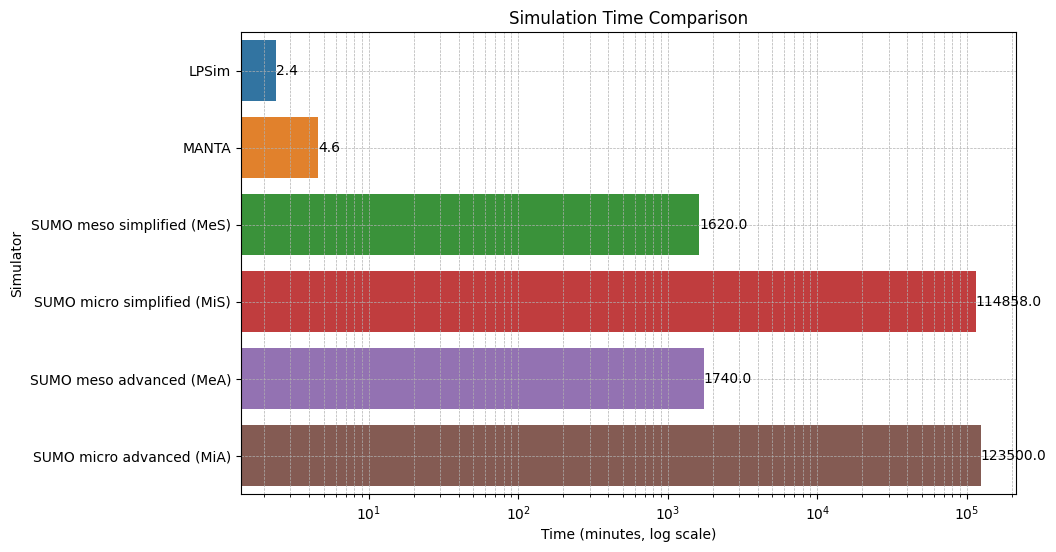}}
\caption{Simulation Time Comparison for Regional Traffic Simulation
}
\label{fig::speed}
\end{figure}

\begin{figure}[!t]
\centerline{\includegraphics[scale=0.6]{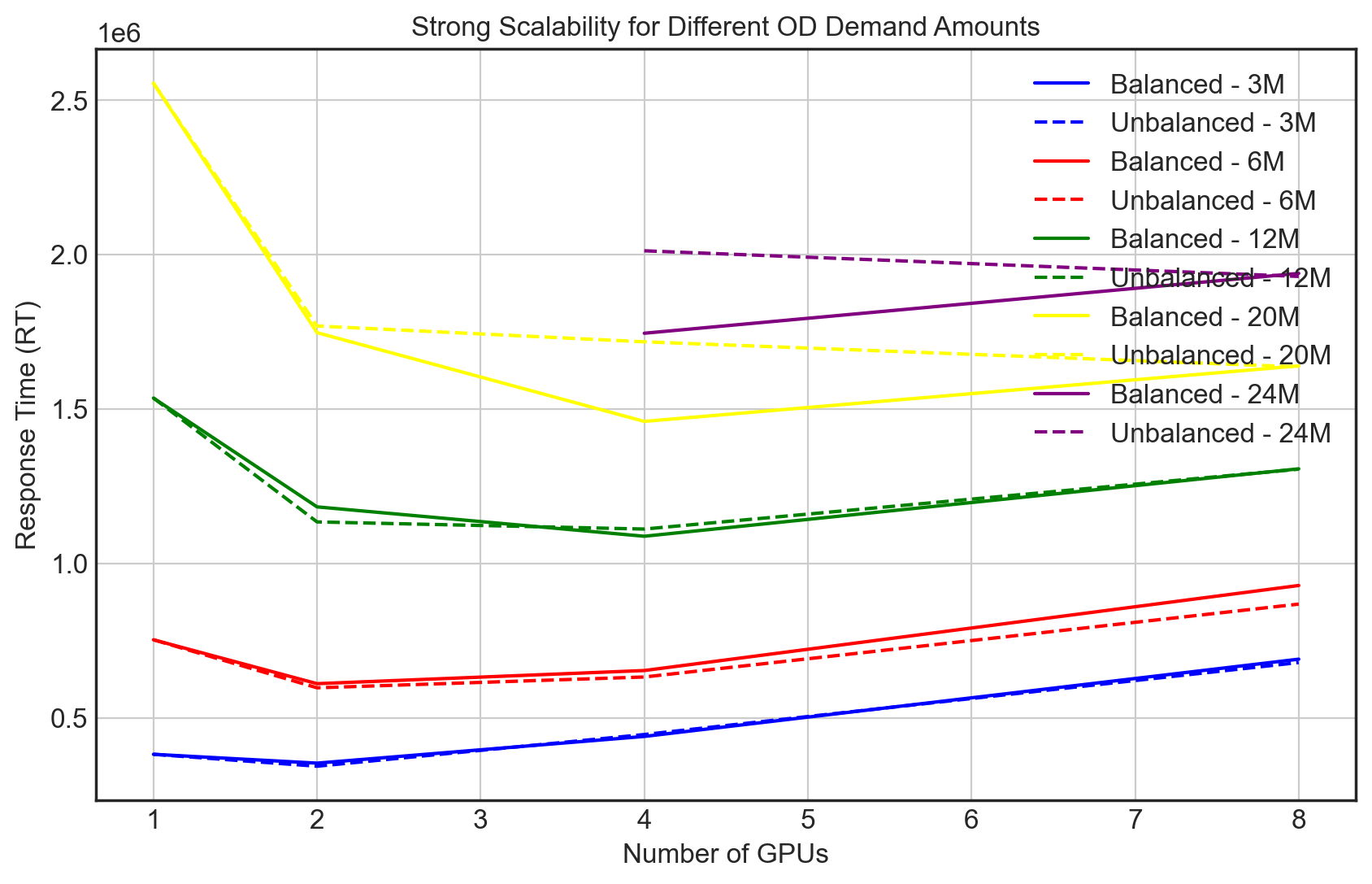}}
\caption{Simulation Time for Different Demands with different number of GPUs}
\label{fig::scale}
\end{figure}

 A prevalent approach in traffic simulation is the use of microscopic simulation \cite{treiber2000microscopic} \cite{maroto2006real}, which refers to a computer-based modeling technique that simulates the behavior and interactions of individual entities at a microscopic level, which models detailed physical dynamics between vehicles. In the traffic field, microscopic simulation takes a granular perspective, focusing on individual vehicle movements, accelerations, and lane changes to provide a realistic representation of traffic flow. Compared to macroscopic simulation and mesoscopic simulation, which focus on complete road flow instead of individual vehicle movement
\cite{helbing1995social}, people often use microscopic simulation on the level of cities and highways \cite{maroto2006real} \cite{du2015microscopic}, which are smaller scale. However, in recent years, the popularity of tens of millions of trips of large-scale microscopic simulation has grown significantly \cite{maciejewski2016large}, and the increasing demand for techniques
% \jwdnote{replace "technique support" by "techniques"?}
to accommodate various new traffic modes. For example, the Urban Air Mobility (UAM) system \cite{muna2021air} has a regional impact, requiring researchers to evaluate it at a regional scale. \cite{yedavalli2021microsimulation} tried to do regional scale simulation with single GPU and can scale up to 3.2M trips. With the same network and demand used by \cite{yedavalli2021microsimulation}, Figure \ref{fig::speed} shows that in the same simulation scale in SF bay area with 3.2M OD trips LPSim stands out to be the fastest simulator of all the Traffic Simulators. Other than the simulation speed, LPSim can scale the simulation up to 24M OD trips and beyond as shown in Figure \ref{fig::scale}, which none of the simulators can achieve yet.

% This includes assessing local congestion caused by UAM and the detailed movement of UAM vehicles to avoid conflicts, all of which must be precisely analyzed through large-scale microscopic simulation.
% \jwdnote{Fix grammar.}
 
Compared to regular-scale microscopic simulation, the aforementioned large-scale microscopic traffic simulation has its challenges. In particular, modeling individual vehicle movements in fine detail, continuously processing, updating the states of numerous individual vehicles, and managing extensive spatiotemporal data in the large-scale microscopic simulation will lead to high computational requirements\cite{algers1997review}. Therefore, for practitioners and researchers, large-scale, time-varying car following model based traffic assignment and propagation call for efficient computation tools and scalable framework.

It is possible to utilize the developed GPU architectures to improve traffic simulation \cite{yedavalli2022microsimulation}. LPSim, presented in this paper, uses GPUs to handle both the spatial and temporal aspects of large-scale microscopic traffic simulations. Spatial aspects include the network partitioning into different GPUs, GPU threads Reallocation between GPUs, and Ghost Zone Creation between GPUs\cite{barcelo2010fundamentals}\cite{boxill2000evaluation}. The temporal dynamics mainly consist of the time-driven propagation and synchronization across time steps of traffic flow for each time step\cite{osorio2015energy}. However, the challenge of GPU memory limitations, with a single GPU's memory capped at 80 GB (for A100)  \cite{choquette2020nvidia} and most GPUs have only 8GB memory \cite{zhang2017gpu}, contrasts with the potential for Central Processing Units (CPUs) to scale up to 768 GB (for AWS) \cite{pelle2019towards}. To overcome this, our framework employs graph partitioning 
% \jwdnote{Change "graph partition" to "graph partitioning", here and later.}
methods to distribute vast amounts of transportation network and vehicle movement data across multiple GPUs. This strategy ensures that the simulation can scale to accommodate large-scale networks without compromising the level of detail or simulation speed. And we used the combination of car following model, lane change model, and gap acceptance model to model traffic propagation which is a common approach used by popular micro simulators, such as SUMO \cite{krajzewicz2002sumo}, Vissim \cite{lownes2006vissim}, Aimsun \cite{casas2010traffic}, we used Intelligent Driver Model (IDM) \cite{kesting2010enhanced} and our framework is adaptable to any other types of traffic propagation models.  We tested our simulator on an Intel(R) Xeon(R) Gold 6326 CPU @ 2.90GHz with a typical day demand of 9008766 trips from 0 AM to 12 PM provided by SFCTA \cite{outwater2006san} on a SF bay area road network with  224,223 nodes and 549,008 edges which finished with 4.49192555556 hours compared to a dual NVIDIA A100-PCIE-40GB GPUs' 0.0398483333 hours simulation time which are around 113 times faster. 
Our main design innovation for the single GPU framework is summarized in Table \ref{table-1}, and for the multi-GPU system is summarized in Table \ref{table-0}. Based on our research, we claim contributions in the following three computational aspects, rather than theoretical transportation models:

\begin{itemize}
    \item  \textbf{Design of the Multi-GPU Transportation Simulation Framework:} We have developed a transportation simulation framework that uses multiple GPUs, structured around the concept of graph partitioning. This design ensures the simulation outputs remain consistent as the number of GPUs increases, thereby enhancing the effectiveness and efficiency of the simulation.
    \begin{itemize}
        \item \textbf{Vectorized Data Storage and Access Implementation:} Our approach incorporates vectorized data storage and access mechanisms that allow for the efficient storage and access of both transportation network data and vehicular movement/speed information within a GPU environment. This innovation facilitates improved data handling and processing speed within the GPU architecture, which are detailed explained in Figure \ref{data}.
    \end{itemize}
    \item \textbf{Scalability Benchmarking}: The scalability of LPSim has been thoroughly evaluated both theoretically and through empirical research on multiple GPUs with Google Cloud Service and generic personal GPUs. Scalability assessments were conducted using data sourced from the San Francisco County Transportation Authority (SFCTA), complemented by performance analysis based on the Bulk Synchronous Parallel Model \cite{gerbessiotis1994direct}, Amdahl's law \cite{hill2008amdahl}, and Roofline Model \cite{williams2009roofline},
    % \jwdnote{Add citation for Roofline}
    LPSim demonstrates remarkable efficiency, when the scale becomes larger and has to be deployed on an Google Cloud instance equipped with V100 GPUs, LPSim can efficiently simulate 9.01 million trips in 21.16 minutes, and gains nearly 1.68 times speedup when increases the number of GPUs from 2 to 4 with balanced partition of the road network. These analyses demonstrate the framework's ability to scale effectively in response to varying computational demands.
    % \item \textbf{Efficient Data Storage Method:} We devise a more efficient method for data storage, significantly increasing GPU utilization rates to 80\%.
    \item \textbf{Trade-off between Multi-GPU Parallel Computing and Communication Time}: Our investigations reveal that while multi-GPU parallel computing can expedite the simulation process, it is also subject to potential slowdowns due to increased communication times among GPUs. Simulation experiments conducted on systems with 2, 4, and 8 GPUs on a gcloud instance with NVIDIA V100 GPUs have yielded the data, as detailed in Table \ref{table:benchmarking}, which is explained further in Section \ref{Performance}. It indicates that the increase in the number of GPUs, coupled with appropriate graph partitioning strategies, can lead to a reduction in total computation time required for simulation.
    
\end{itemize}

% \begin{table}[]
% \setlength\tabcolsep{3pt}
% 	\caption{SINGLE GPU COMPUTATION DESIGN.\ref{airnetwork}}
% 	\label{table-1} 
% 	{\small 	\begin{center}
% 			\begin{tabular}{ccccc}
% 				\toprule
% 				{\bf Traffic Data } & {\bf Structure } &  {\bf Architecture Design  } \\
% 				\midrule
% 				Network &  Large scale data \footnote{The network contains half a million edges} & Contiguous memory blocks\\ 
% 				\midrule 
% 					Vehicle info& Spatiotemporal data  &  GPU threads $\rightleftharpoons$ vehicle\\
% 				\midrule
% 				Agent status Eq. \ref{eq:vehicle_dynamics}& Markov Update &  GPU SIMD  \\
% 				\bottomrule
% 			\end{tabular}
% 	\end{center} }
%  \vspace{-0.3cm}
% \end{table}
\begin{table*}[ht]
    \centering % This centers the table
    \begin{threeparttable}
    \caption{SINGLE GPU COMPUTATION DESIGN.}
    \label{table-1}
    \setlength\tabcolsep{3pt}
    \footnotesize

    \begin{tabular}{p{0.3\linewidth} p{0.65\linewidth}}
        \toprule
        {\bf Single GPU Challenge} & {\bf Solution} \\
        \midrule
        Traffic Atlas & Simultaneous Representation of Road Network, Vehicle Speeds, and Locations inside GPU Memory (see Fig \ref{data})\\
        \midrule
        Vehicle States & GPU Threads \cite{gupta2012study}$\rightleftharpoons$ Vehicles\\
        \midrule
        Traffic Atlas and Vehicle States' Markov Update & GPU single instruction multiple data (SIMD) \cite{yilmazer2014scalar} \\
        \bottomrule
    \end{tabular}
    \end{threeparttable}
\end{table*}

\begin{table*}[ht]
\setlength\tabcolsep{4pt}
	\caption{MULTI-GPU COMPUTATION DESIGN.}
	\label{table-0}
	{\small 	\begin{center}
			\begin{tabular}{ccccc}
				\toprule
				{\bf Multi-GPU Challenge } & {\bf Solution  }\\
                \midrule
                Multi-GPU Load Balancing & Graph Partitioning Fig \ref{fig:framework-partition} \\
				\midrule
				Multi-GPU Communication &  Ghost Zone Design Fig \ref{fig_ghost}\\ 
				\midrule 
				Multi-GPU Vehicle Threads Dynamic Reallocation & GPU Device based Vector Fig \ref{vehicles_storage_array_vector} \\
				\bottomrule
			\end{tabular}
	\end{center} }
 \vspace{-0.3cm}
\end{table*}

The structure of this article is laid out as follows: Section \ref{Literature review} provides some related work of this paper. Section \ref{method} dives into the specifics of our proposed methodology. Following this, Section \ref{results} details the experiments conducted and discusses their outcomes. The performance benchmarking of our approach is then thoroughly examined in Section \ref{Performance}. The article is brought to a conclusion in Section \ref{Conclusion}.
% \begin{table}[ht]
% \centering
% \setlength\tabcolsep{4pt}
% \caption{Simulation Time /ms}
% \label{tab:numerical_intro}
% \begin{tabular}{|c|c|c|c|}
% \hline
% GPU Numbers         & 2        & 4       & 8       \\ \hline
% Random Partition    &  Dumped        &    Dumped     &  Dumped       \\ \hline
% Balanced Partition  & 2498466  & 1483472 & 1269893 \\ \hline
% Unbalance Partition & 2014554 & 1679861 & 1783668 \\ \hline
% \end{tabular}
% \end{table}

\section{Related Work}
\label{Literature review}
\subsection{Computational Perspective - From CPU to GPU to multiple GPU}
 While CPUs have traditionally served as the backbone of general-purpose computing, the emergence of GPUs has triggered a change in processing capabilities. This shift is rooted in the parallel architecture of GPUs, which allows them to simultaneously handle a multitude of tasks. Especially, GPUs excel at SIMD processing \cite{bhandarkar1996parallel}\cite{franchetti2005efficient}, where a single instruction is executed on multiple data points simultaneously. This is beneficial for tasks involving repetitive and parallelizable computations, such as those found in graphics rendering and scientific simulations. In addition, GPUs feature a high-bandwidth memory hierarchy that allows quick access to large datasets \cite{mei2014benchmarking} 
 \cite{mei2016dissecting}. This is crucial for applications like gaming and data-intensive computations, where rapid access to a vast amount of data is essential for better performance \cite{kim2011achieving}.
 As the demands for faster and more efficient processing continue to surge, harnessing the collective power of multiple GPUs is receiving more and more attention. The integration of multiple GPUs represents a substantial advancement, where the collective output transcends the capabilities of individual units with combined memory to enable the scaling to bigger scenarios. This transition not only increases available computational resources but also accelerates the execution of tasks through the synergistic effect to conquer memory constraints \cite{schaa2009exploring} \cite{stuart2011multi}. However, the integration of multiple GPUs for parallel computing presents a unique set of challenges, which arise from the need to efficiently distribute and synchronize tasks between multiple processors, manage data transfer between GPUs, and address potential bottlenecks of communication and data races. Additionally, software scalability and compatibility become crucial factors, as not all applications can seamlessly leverage the parallel processing capabilities of multiple GPUs \cite{xiao2010inter}. For researchers, navigating the challenges mentioned above is essential for unlocking the full potential of multi-GPU systems and maximizing computational efficiency. 
% GPU-based traffic simulation Framework
\subsection{Traffic Perspective - Simulation Implementation and Framework Design}
The aforementioned review of GPUs and multi-GPU computing highlights both advantages and challenges. For general computation missions, addressing such challenges will be difficult and there is no unified framework using multi-GPU computing up to now. In specific research domains, researchers have done works utilizing multi-GPUs in the area of numerical linear algebra \cite{agullo2011qr}, graph analytics \cite{pan2017multi}, optimization algorithm solvers \cite{ament2010parallel}, and so on. 
As far as we know, the use of multiple GPU computations in transportation modeling and simulation tools has not yet achieved widespread adoption. We give a review of the previous simulation strategies in the traffic area first. 

Historically, discrete-event simulation and network flow simulation were the standard techniques used in transportation simulation \cite{borgatti2005centrality}. Their capabilities are further examined in relation to two key functionalities of any transportation simulator: traffic operation \cite{jansen2004operational} and dynamic routing \cite{savelsbergh1998drive}. With the advent of multi-modal simulators and the concurrent challenges they posed, agent-based simulation was introduced \cite{railsback2006agent}. 
This approach has a significant impact on the execution speed of the simulations. However, integration of these functions tends to complicate the model, often resulting in slower processing speeds. As functionality becomes more and more complicated, the datasets also become larger and larger.

\begin{figure}[!htbp]
\centering
\includegraphics[scale = 0.7]{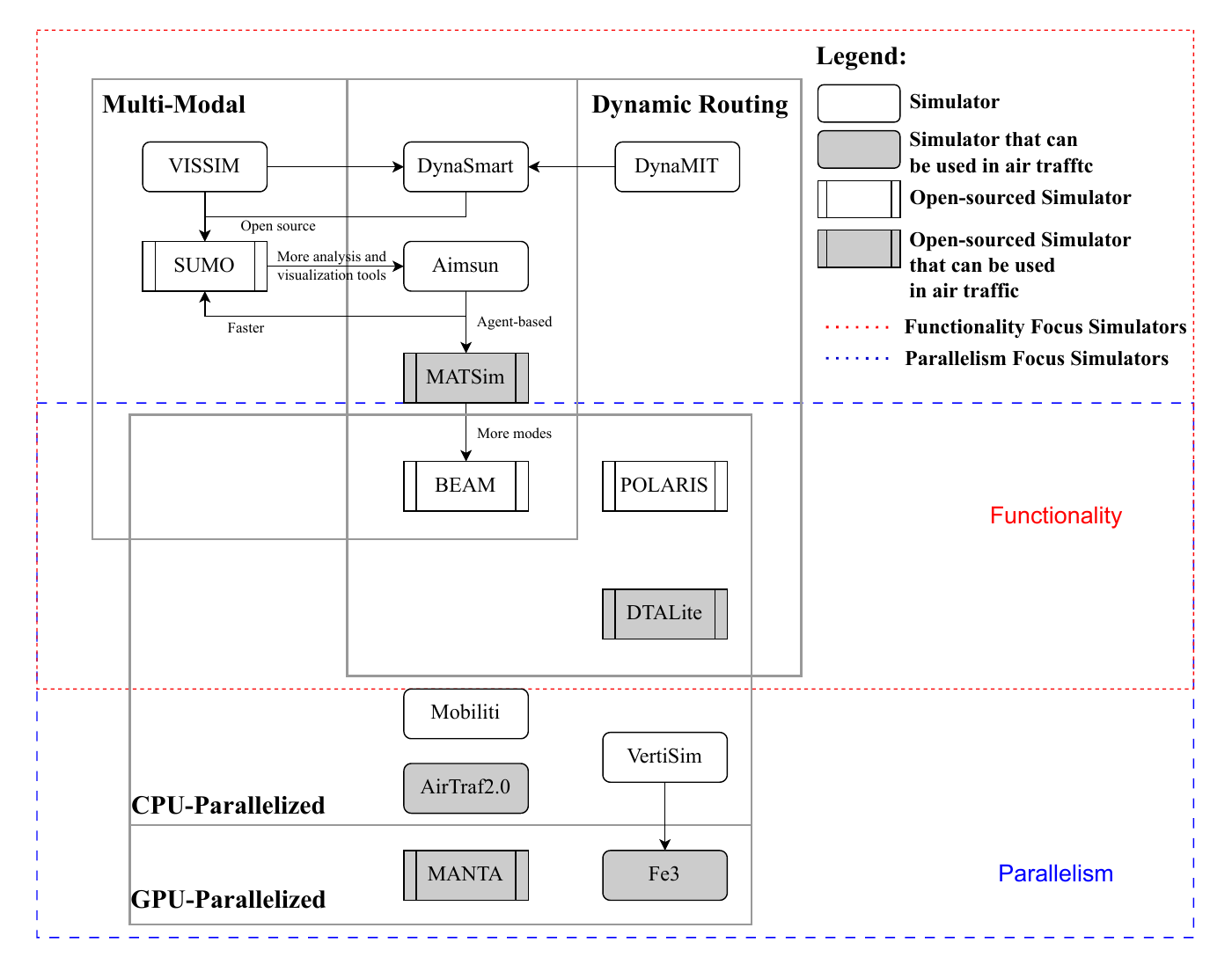}
\caption{Simulation Evolution Progress Representation \cite{lownes2006vissim, mahmassani2002dynasmart, ben2002real, zhao2017dynamic, casas2010traffic, ronaldo2012comparison, behrisch2011sumo, w2016multi, sheppard2017modeling, sheppard2016cost, chan2018mobiliti, yedavalli2021microsimulation}
% \jwdnote{Changed "paralleled" to "parallelized" (2x)}
}
% \todo{put all the references in the caption here @Yuhan}
% \todo{@Xuan}
\label{simulation_survey}
\end{figure}

One solution in transportation to effectively manage large datasets and complicated functions is the use of supercomputers, as Mobiliti \cite{chan2018mobiliti} did. Another direction is harnessing the parallel structure of CPU/GPU, as MANTA \cite{yedavalli2021microsimulation} uses one GPU in parallel with the CPU 
% \jwdnote{Do you mean in parallel with the CPU?} 
to greatly enhance simulation speed. Some traffic simulation software projects, such as BEAM \cite{sheppard2017modeling} and POLARIS \cite{auld2016polaris}, have also enabled the use of CPU-based parallelism with mesoscopic simulation. The evolution of popular simulations is summarized in Figure \ref{simulation_survey}: We illustrate the simulators' capabilities with dashed red lines, such as different modes of transportation and dynamic routing of people's route choices.
% \jwdnote{of what?}. 
The dashed blue boundary indicates the growing emphasis on parallelism in recent studies, with some simulators like BEAM, POLARIS, and DTALite focusing on both functionality and parallelism. The arrows depict the evolutionary trajectory of these simulators. For instance, VISSIM, DynaSmart, being proprietary, have led researchers to consider open-source alternatives like SUMO. Aimsun, on the other hand, was developed in response to perceived gaps in SUMO's analysis and visualization capabilities, despite SUMO's faster simulation speeds. Moreover, the development of Multi-Agent Transport Simulation (MATSim) signifies a pivotal transition towards agent-based modeling in the transportation sector, diverging from traditional trip-based approaches. This evolution reflects an increasing demand for simulations that can more accurately represent individual behaviors and interactions within transportation networks. Expanding upon MATSim's foundation, Behavior, Energy, Autonomy, and Mobility (BEAM) introduces enhancements that enable the simulation of electric vehicles, shared mobility services (e.g., bicycles), and the utilization of CPU-based parallelism. The gray boxes signify simulators capable of air traffic simulation, a complex 3D challenge distinct from ground traffic. However, the theory of parallelizable GPU-based traffic simulation and the approach of using multiple GPUs to not only utilize the computation power but also use the added GPU memory to simulate bigger scenarios faster have not been investigated yet.

% Therefore, the combination of multiple GPUs for traffic simulation can not only enhance the scale of the propogation of the vehicles but also the computation time of it will be shorten. Previous researches also proved that the idea is very encouraging and promising. Yedavalli \textit{et al.} \cite{yedavalli2021microsimulation} simulated the peak hour trips within a 
 
%Although researchers are exploring the research in parallel computing, the full potential of parallel computing has not yet been exploited.

% \todo{Adding the statements here, simulation focuses on functionality, and then parallelism etc but it's not successful because of balabala, and then refer to Figure \ref{simulation_survey}. @Yuhan}

\section{Theoretical Analysis of GPU-based traffic simulation}
\label{method}
\subsection{Basic Components}
\label{lab:basic_components}
% Multiple and single GPU both have (traffic simulation vehicle dynamic funcitons, traffic atlas, multi modal applications)

\begin{table*}[!htbp]
\centering
\small % Reduce font size
\caption{List of Notations for Basic Components \ref{lab:basic_components}}
\label{table_notation}
\renewcommand{\arraystretch}{1.2} % Adjusts the row height
\begin{tabular}{|c|p{6cm}|c|p{6cm}|} % Adjust column width
\hline
\textbf{Symbol} & \multicolumn{1}{c|}{\textbf{Description}} & \textbf{Symbol} & \multicolumn{1}{c|}{\textbf{Description}} \\
\hline
$T$ & Timestep of the simulation & $S$ & Spatial resolution each byte of memory represents \\
\hline
$l_{\text{min}}$ & Minimum length of the link of the road network & $v_i(k)$ & Speed of vehicle $i$ at time step $k$\\
\hline
$\Delta v$ & The velocity difference between a vehicle and the one ahead of it & $\l_i(k)$ & The location of vehicle $i$ at time step $k$ \\
\hline

$m_i(k)$ & Probability of a mandatory lane change for vehicle $i$ at time step k& $x_i(k)$ & Distance of vehicle $i$ at time step k to an exit or intersection \\
\hline
$x_0$ & Distance of a critical location to the exit or intersection & $g_{\text{lead}}(k)$ & Critical lead gap for a lane change at time step $k$\\
\hline
$g_{\text{lag}}(k)$ & Critical lag gap for a lane change at time step $k$& $g_a$ & Desired lead gap for a lane change \\
\hline
$g_b$ & Desired lag gap for a lane change & $v_a$ & Speed of the lead vehicle \\
\hline
$v_b$ & Speed of the lag vehicle & $\alpha_i$ & Anticipation time of vehicle $i$ \\
\hline
$\alpha_a$ & Anticipation time of the lead vehicle & $\alpha_b$ & Anticipation time of the lag vehicle \\
\hline
$\epsilon_a$ & Random component for lead gap & $\epsilon_b$ & Random component for lag gap \\
\hline
$\dot{v}$ & Current acceleration of the vehicle & $a$ & Acceleration potential of the vehicle \\
\hline
$v$ & Current speed of the vehicle & $v_o$ & Speed limit of the edge \\
\hline
$\delta$ & Acceleration exponent & $s$ & Gap between the vehicle and the leading vehicle \\
\hline
$s_0$ & Minimum spacing between vehicles at a standstill & $b$ & Braking deceleration of the vehicle \\
\hline
$T$ & Desired time headway & $v_{\text{free}}$ &  Free Flow Speed \\
\hline
$f_i$ & The function abstraction of IDM car following, lane change, and gap acceptance & $v_f(k)$ & Front vehicle's speed at time step k\\
\hline
$l_f(k)$ & The location of front vehicle at time step k & &\\
\hline
\end{tabular}
\end{table*}
In this part, we introduce the vehicle dynamics used for microsimulation initially developed in \cite{yedavalli2021microsimulation}. There are three key components: car following model, lane change process, and gap acceptance. 

We further present a theory of parallelizable simulation that aligns with the GPU's SIMD architecture as shown in Eq. \ref{eq:vehicle_dynamics}. This includes a detailed explanation of how memory is utilized not only to represent the road network but also to depict the occupancy and speed of the vehicles. Lastly, we consolidate our methodology into a comprehensive algorithm, outlined as Algorithm \ref{alg:vehicle_movement}. Readers can refer to table \ref{table_notation} for notation in this part.

\textit{Vehicle Propagation: }First, the dynamics of vehicles within our simulation is governed by the car following model as described in Equation \eqref{eq:velocity}, which models vehicle acceleration to update vehicle locations at each time step. Second, the lane change process is encapsulated by Equation \eqref{eq:lane_change}. For the vehicle, when the distance to an exit falls below a threshold distance $x_0$, it is triggered to make a mandatory lane change and the probability of such a change increases as the vehicle approaches the exit. Once a vehicle inquires for a lane change, it must assess the feasibility of this maneuver by assessing the gaps with both the leading and the lagging vehicles. This assessment is carried out according to Equation \eqref{eq:glead_glag}. For LPSim, we adapt the IDM model as described in \cite{albeaik2022limitations, iqbal2014development, choudhury2007modeling}

% Since \cite{yedavalli2021designing} has validated their results with Uber movement project, our first step will be comparing the results from our model with theirs to confirm the correctness of our results. and we'll further introduce SFCTA data to understand the performance of our model and push further the boundary of our model.

\begin{equation} \label{eq:velocity}
v_i(k+1)= f_{a, v_0, \delta, s_0, T, b }(v_i(k), v_f(k), l_f(k), l_i(k)) \tag{Car Following}
\end{equation}

\begin{equation} \label{eq:lane_change}
m_i(k+1)= f_{x_0}(x_i(k))
\tag{Lane Change}
\end{equation}

\begin{align}
\label{eq:glead_glag}
g_{\text{lead}(k+1)} &= f_{g_a, \alpha_a, \alpha_i, v_a, \epsilon_a}(v_i(k), l_{i-n}(k), \ldots, l_{i+n}(k))
\tag{Gap Acceptance}\\
\label{eq:glead_glag}
g_{\text{lag}(k+1)} &= f_{g_b, \alpha_b, \alpha_i, v_b, \epsilon_b}(v_i(k), l_{i-n}(k), \ldots, l_{i+n}(k))
\tag{Gap Acceptance}
\end{align}

Incorporating the components previously outlined, the traffic simulation process for each subsequent time step is effectively encapsulated by Equation \eqref{eq:vehicle_dynamics}. This equation dictates that the position of vehicle $i$ at time step $k+1$ is determined by a specific set of factors: the vehicle's position and velocity at the preceding time step $k$, the positions of surrounding vehicles at time step $k$, and the speed of the vehicle directly ahead at time step $k$. The computation for each vehicle at time step $k+1$ is conducted independently, relying solely on static data from the previous time step $k$. This approach aligns perfectly with the SIMD architecture of GPUs, enabling efficient parallel processing of traffic simulations.

\begin{equation}
l_i(k+1) = f_i( \underbrace{l_i(k), v_i(k)}_{\text{Current Vehicle}}, \underbrace{l_{i-n}(k), \ldots, l_{i+n}(k)}_{\text{Surrounding Vehicles}}, \underbrace{l_{f}(k),v_{f}(k)}_{\text{Front Vehicle}} )
\label{eq:vehicle_dynamics}
\end{equation}

% Where $i$ denotes the $i$-th vehicle and $k$ the current time step. The symbol $l$ represents the location, while $u$ denotes the dynamics of vehicles. Therefore, $x_{i-n}, \ldots, l_{i+n}$ represents the locations of adjacent vehicles.
Since $l_i(k+1)$ is only dependent on the state in the previous time step $k$, a parallel computation of all vehicle states in time step $k+1$ can be implemented using multiple threads simultaneously. The vehicle propagation process for each vehicle is shown in Algorithm \ref{alg:vehicle_movement}.

\textbf{Remark}: 
\begin{itemize}
    \item \textbf{Network Information Representation: } One byte can only be occupied by one vehicle, and we are not simulating the case that vehicles crash with each other. 
    \item \textbf{Intersection Modeling: } When vehicles approach, they will wait before the intersection and check the downstream road segment. If the downstream road segment is full, then the vehicles will be waiting at the current road segment.
    \item \textbf{Switch from one road to another road segment: } When a vehicle transitions from one road segment to another, we employ atomic operations to ensure data integrity. Specifically, if two vehicles, A and B, attempt to move to road segment C simultaneously, atomic operations guarantee that only one vehicle's thread successfully writes to segment C. This prevents the issue of both vehicles occupying the same memory address, which could otherwise result in vehicle data conflicts and potential loss of vehicle information.
\end{itemize}

% \jwdnote{Say how this prevents 2 vehicles from occupying the same space, or if crashes can be simulated.}

As shown in Figure \ref{data}, the data stores inside GPU memory contains three components:
\begin{itemize}
    \item \textbf{Lane Map: }The whole road network will be represented in GPU memory with Char type in C++ which can represent Numbers from 0 to 255. We use it in the following way:
\begin{itemize}
    \item \textbf{1 Byte in Memory = 1 Meter in Road Segments:} This means each byte corresponds to one meter of road in the simulation.
    \item \textbf{Value 255 = Not Occupied:} Indicates that the byte is not occupied by any vehicle.
    \item \textbf{Values 0-254 = Occupied by Vehicle:} Represents that the byte is occupied by a vehicle, with the value indicating the vehicle's speed in meters per second.
\end{itemize}

\textbf{Limitation: }Using the Char type means that the maximum value we can represent is 255. Therefore, if a vehicle's speed exceeds 254 meters per second, this framework cannot accurately represent it. Furthermore, since each byte corresponds to one meter, the precision of vehicle placement and road representation is limited to 1 meter.

We use an adjacency list to map the directed connectivity of nodes, with list entries pointing to arrays that describe road segments (e.g., a 1D array for a 4-lane, 8-meter road segment will be the dimension of 1 X (4*8), resulting in dimensions of 1 X 32 as shown in Fig \ref{data}.) Using Char Type in C++ to indicate road status (occupied, empty, and vehicle speed, etc.), the real road segment is mapped in a 1D array \textbf{lane map} organizing lanes in order.
    \item \textbf{Network Data Information: }For each Edge from input, we extract and store the number of lanes, its index on lane map, upstream intersection, and downstream intersection information.
    \item \textbf{Vehicle Data Information: }For each vehicle, the route path and the first edge that it is going to traverse are from the input demand data after the routing. When the simulation is running, we keep a record and update each vehicle's previous edge, current edge, next edge, and position on the current edge.
\end{itemize}
With three components above, when we are calculating each vehicle's movement, we calculate vehicle locations at the lane map according to the edge's index on the lane map and its position on the current edge to update the lane map

% \begin{figure*}[!htbp]
% \centering
% \includegraphics[scale = 0.45]{image/one_gpu_road_1.pdf}
% \caption{One GPU representation of Road Network and Vehicles
% % \jwdnote{What do the two dimensions of the
% % array mean? How does it relate to the description of what 1 Byte represents? The text refers to a
% % 10 meter road segment, which should be 10 bytes,
% % but this array is 8x8.}
% }
% \label{airnetwork}
% \end{figure*} 

\begin{sidewaysfigure*}
\centering
\includegraphics[scale=0.43]{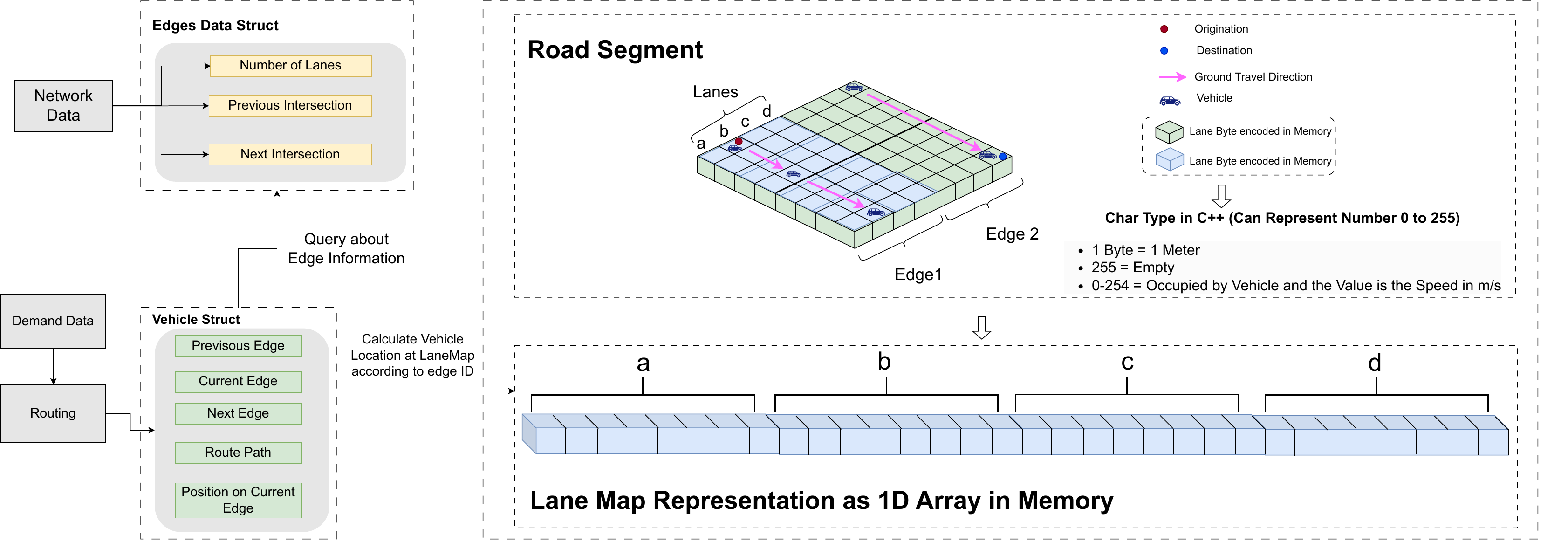}
\caption{Data Movement Representation inside GPU}
\label{data}
\end{sidewaysfigure*}

\subsection{Architecture of single GPU-based traffic simulation}
\label{one_gpu}
% \textcolor{red}{Jiaying}

% \subsubsection{Data Structures and Memory Management}

% Effective data organization and management are vital for optimizing performance in GPU-based traffic simulations. In our simulation system, key entities like intersections, traffic lights, road networks incorporating cars' speeds, and vehicles detailed with attributes such as velocity, acceleration, and location, are all methodically structured in arrays stored in contiguous memory blocks within the GPU's global memory. Due to the efficient element access of arrays, the arrangement above perfectly aligns with the GPU's parallel processing architecture, allowing each thread on the GPU to read and update the data efficiently and simultaneously. As a result, it will boost the capabilities for parallel simulation execution.

% Pretty sure you guys don't want to say "Verti-Pot"

% \todo{Adding other modes like transit, bike, pedestrians etc into this figure @Yuhan}

\subsubsection{GPU Kernel Design for Vehicle and Traffic Management}

In our GPU-based traffic model, the kernel design is a cornerstone of ensuring that the traffic management and vehicle state updates are performed efficiently. Each GPU thread is responsible for updating the state of an individual vehicle and its interaction with the traffic network per simulation time step.

\begin{algorithm}[!htb]
\caption{Vehicle Propagation Algorithm}
\label{alg:vehicle_movement}
\textbf{State}: $\mathcal{R}$ (Road Status), $t_{depart}$ (Departure Time of current vehicle), $\mathcal{V}$ (Vehicle Type), $e_{current}$ (Current Edge), $v$ (Speed), $\mathcal{I}$ (Intersection Status)\\
\textbf{Parameter}: $\Delta t$ (Time Step), $v_{free}$ (Free Flow Speed), $\Theta_{IDM}$ (IDM Parameters $\left(a, \delta, b, s_0\right)$ as explained in Table \ref{lab:basic_components})\\
\textbf{Output}: $\mathcal{P}$ (Vehicle Position), $v$ (Speed), $a$ (Acceleration)
\begin{algorithmic}[1] %[1] enables line numbers
\While{Vehicle is active}
    \If{$\mathcal{R}$ is occupied $\lor$ Current Time $< t_{depart}$}
        \State Wait
    \Else
        \State Start moving, we can write vehicle status on $\mathcal{R}$ and update $\mathcal{P}$ accordingly.
        \If{Moving on a New Edge}
            \State $e_{current} \gets$ New Edge ID
            \State $t_{start} \gets$ Current Time
            \State Record $\mathcal{V}$ type
            \State $d_{front} \gets 2 \cdot \Delta t \cdot v$ (because maximum distance a vehicle can move within $\Delta t$ is $v \Delta t + \frac{1}{2}a\Delta t^2 = v \Delta t + \frac{1}{2}v\Delta t = \frac{3}{2}v\Delta t$, we check more than the maximum distance to avoid collision)
        \EndIf

                \If{Front car within $d_{front}$}
            \State Update $\mathcal{P}, a, v$ using $\Theta_{IDM}$
        \Else
            \State $v \gets v_{free}$
        \EndIf
        \If{Intersection reached}
            \State Proceed according to $\mathcal{I}$'s signal controls
        \Else
            \State Evaluate lane change with Equation \eqref{eq:lane_change}
            \If{Changing lane}
            \If{Gap acceptance check with Equation \eqref{eq:glead_glag}}
                \State Change lane
            \EndIf
            \EndIf
        \EndIf
    \EndIf
\EndWhile
\State \textbf{return} $\mathcal{R}$ , $\mathcal{P}, v, a$
\end{algorithmic}
\end{algorithm}

In detail, as shown in the Vehicle Propagation Algorithm \ref{alg:vehicle_movement}, the $\mathcal{R}$ Road Status array represents the occupancy status or speed of vehicles on the road, while $\mathcal{I}$ (Intersection Status) denotes the graph nodes. The process begins with each thread checking the road status and vehicle's departure time to decide whether to depart or wait. Once the vehicle is active, the thread orchestrates the vehicle's progression, revising its position and velocity in response to the immediate road conditions, such as the presence and distance of preceding vehicles. Additionally, the thread evaluates the vehicle’s interaction with intersections and potential lane changes.

\subsubsection{Limitations of Single GPU Architecture and Evolution to Multi-GPU}

The limitations of single GPU architectures are notable, particularly in scalability and memory. As traffic simulations grow more complex, a single GPU's finite cores and bandwidth may lead to longer simulation times. Additionally, the memory limit can constrain the scope of simulatable traffic scenarios. 

Multi-GPU framework is an extension of the single GPU architecture, scaling up by distributing the workload across multiple GPUs. This approach inherits the core principles and functionalities of single GPU systems, such as kernel functions and parallel processing, while addressing the performance and memory limitations by enabling larger and more complex traffic simulations. 

\subsection{Design of multiple GPU-based simulation framework}

% \textcolor{red}{Yuhan/Code part Design -- Check with correctness with Jiaying}

In scenarios where only a single GPU can be utilized, the entire graph comprising nodes (representing intersections, points of interest, etc.) and edges (depicting the roads or paths between nodes) is stored in one GPU. When multiple GPUs are engaged, the graph is partitioned across these GPUs. This partitioning necessitates an approach to handling the vehicle movements across multiple GPUs. To facilitate this, the graph is divided, and the nodes are distributed across multiple GPUs. Edges that connect nodes situated on different GPUs are placed within a so-called 'ghost zone', which acts as a replicated buffer area that possesses the same information across the boundaries of adjacent GPUs, ensuring consistency and continuity of information on the network.

In the process of graph partitioning, data are categorized into two types, global data and local data. Global data, which includes the node partitioning scheme (indicating the allocation of specific nodes to particular GPUs) and the complete route of each vehicle, is accessible across all GPUs. Local data such as individual vehicle details in the simulation, the layout of the lanes, intersections, and traffic signal statuses, are stored distinctly within a local GPU.

\begin{figure}[!htbp]
\centerline{\includegraphics[width=\columnwidth]{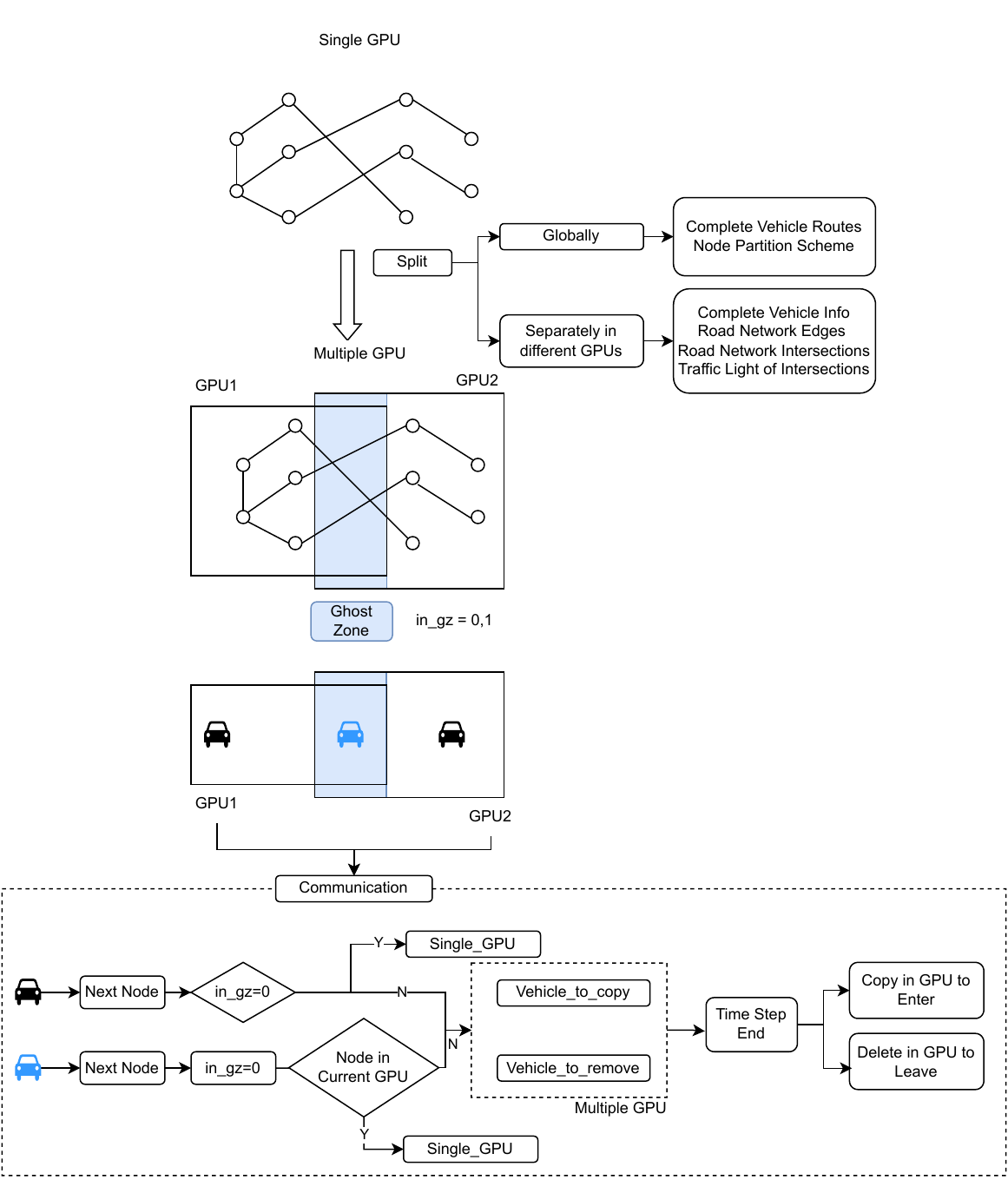}}
\caption{The Framework of the Multi-GPU based traffic simulation
}
\label{fig_ghost}
\end{figure}

The propagation of vehicles and the complete road network across multiple GPUs calls for inter-GPU communications to synchronize state and share data. When a vehicle is not located in the ghost zone, the simulator checks whether its next node belongs to the ghost zone. If not, the simulation continues to a single-GPU scenario, eliminating the need for cross-GPU computations. Conversely, if the vehicle is about to enter the ghost zone, it is duplicated in the ghost zone of the destination GPU. If a vehicle is already in a ghost zone, it is assured that its next node is out of the ghost zone which means after the current edge traversal, the vehicle will be on the next GPU and leave the original GPU. At this juncture, it is determined whether the node resides within the current GPU's domain. If it does, the situation reverts to a single-GPU model as described in Session \ref{one_gpu} again. If the node is outside the current GPU, the vehicle is marked for removal, and subsequently eliminated post the completion of the simulation timestep. The whole procedure is summarized in Figure \ref{fig_ghost}, intersections within the road network are represented by circles, and we use an adjacency list to illustrate the directed relationships between these intersections. The entries in the list link to the arrays that define road segments. For example, a 4-lane 8-meter road segment is depicted as a 1D matrix with a dimension of 1 x 32 as shown in Fig. \ref{data}, showing how real road segments, represented by lines in Fig. \ref{fig_ghost}, are systematically translated into 1D matrices to sequentially organize lanes. In the process of selecting ghost zones, we replicate the entire edge (represented as a 1D array) across both GPUs and the vehicles on the ghost zone will be calculated simultaneously within two GPUs. To ensure exclusive byte allocation to each vehicle, 
% \jwdnote{So every vehicle is treated as being 2 meters long? No trucks :) ?} -- Re: yes, for now we only consider cars, but this framework is extendable and we'll incorporate other modes in the future.
we transfer vehicles one at a time from one edge to another if the first-byte memory of the downstream edge is not occupied.

 By distributing the computation onto different GPUs, our approach allows for scalability and efficient processing of complex simulations that would otherwise be computationally impractical on a single GPU system. We'll explain the detailed theory and implementation of the above framework in the following subsections.
 
% \textcolor{red}{Chonghe/Graph Partition}
\subsubsection{Graph Partitioning}
\begin{figure}[!htbp]
    \centering
    \includegraphics[width=1\columnwidth]{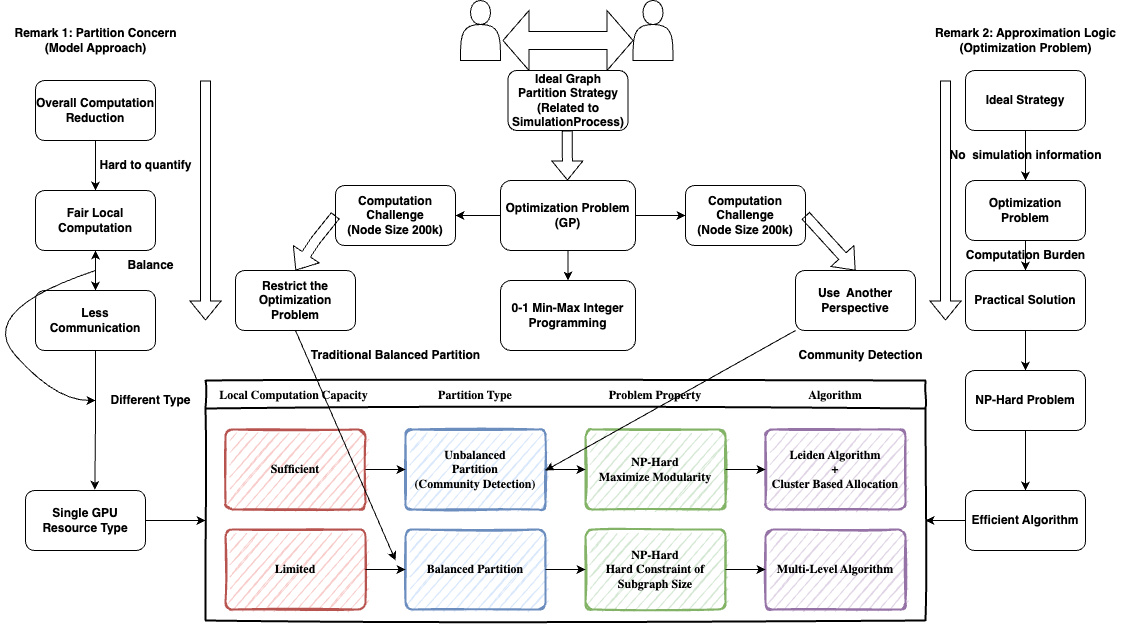}
    \caption{Framework of the graph partitioning strategy}
    \label{fig:framework-partition}
\end{figure}

In large-scale multi-GPU parallel simulations, it is essential to distribute graph data across multiple GPUs to share computational resources. We aim for an efficient allocation of extensive graph data, managing computational resources while minimizing inter-GPU communication. In this part, we first state our problem and propose an approximate optimization formulation of the graph partitioning task. Since it is difficult to solve for our network with $200k$ node size, we introduce two applicable graph partitioning strategies to solve the problem. The choice of two graph partitioning methods is closely related to the available computational resources of the single GPU, and the inherent approximation idea of the two methods comes from different perspectives. The framework is summarized as shown in Figure \ref{fig:framework-partition}, which illustrates that when we have insufficient computational resources, which means that we are creating many more threads than CUDA cores, it leads to a computationally bound scenario. We propose using balanced partitioning to allocate computation evenly across GPUs. In contrast, with ample computational power, the challenge shifts to minimizing communication overhead, necessitating unbalanced partitioning. Our framework consists of three parts: identifying the difficulties in the partition problem, simplifying it to an optimization formulation, and using efficient algorithms to solve the approximation problem. The left side of Fig.~\ref{fig:framework-partition} is about modeling the problem, the right side is about how to formulate the problem and solve it, and the middle part is about the detailed components of our solutions.

(\textbf{Optimization Problem of Graph Partitioning}) The problem of the GPU partitioning is related to the process of simulation. An ideal partition captures the communications of the multi-GPU system at every time step and seeks to minimize them. However, it relies on the propagation of the simulator and is impossible to achieve before the simulation starts. Below, we formally expressed our optimization formulation based on the route information in the studied time. This optimization problem aims to minimize the calculation in the system under the partition scheme. The optimization is summarized below:

% \begin{equation}
%  \begin{aligned}
%     &\min \max \sum\limits_{i,j \in I} A_{ij}\left(\tau_{com}|x_{ik}-x_{jk}|+\tau_{cal}\left[x_{ik}-x_{jk}-1\right]^+\right)\\
%     \text{s.t.} & \sum\limits_i x_{ik} \le \overline{l}, \quad\forall k \in K\\ 
%     & \sum\limits_k x_{ik} = 1,\quad \forall i \in I \\
%     & x_{ik}  = \{0,1\}, \quad \forall i \in I, \forall k \in K 
% \end{aligned}   
% \label{op:partiion}
%     \tag{GP}
% \end{equation}

\begin{equation}
 \begin{aligned}
    \min & \quad s \\
    \text{s.t.} & A_{ij}\cdot\tau_{com}|x_{ik}-x_{jk}|\leq s, \forall i,j,k\\
    & \sum_i x_{ik} \le \overline{l}, \ \forall k \in K \\
    & \sum_k x_{ik} = 1,\ \forall i \in I\\
    & x_{ik}  \in \{0,1\}, \ \forall i \in I, \forall k \in K 
\end{aligned}   
\label{op:partition}
    \tag{GP}
\end{equation}

In problem \ref{op:partition}, we use the following notation:
$A_{ij}$ represents the mean number of vehicles that will be travelling from node $i$ to node $j$ within the studied time period.
$x_{ik}$ is an indicator variable that takes the value $1$ if node $i$ is in partition $k$, $0$ otherwise.
$\tau_{com}$ is the average time needed to communicate then calculate a vehicle between GPUs.
$\tau_{cal}$ is the average time needed to calculate a vehicle on a GPU. The set $I$ and $K$ are the sets of graph nodes and the partition indices.

The optimization problem \eqref{op:partition} is $0-1$ min-max integer programming, for which the computational burden will be unaffordable when the graph node size grows to $200k$. This computation issue triggered us to approximate the optimization problem \eqref{op:partition} in different ways.

(\textbf{Practical Balanced Partition - convert worst case to average}) We note that optimization problem \ref{op:partition} is optimizing the worse communication case. If we relax the worst case scenario to the average case and add a balanced subgraph size restriction, the problem will be similar to the well-known balanced graph partitioning problem \cite{bulucc2016recent}. 

For a specific $(k, 1+\varepsilon)$ balanced partition problem, it minimizes the cut, i.e. the total weight of the edges crossing the partitions of $G$ into $k$ components with the size of each component satisfies the constraint below.
\begin{equation*}
 (1-\varepsilon)\left\lceil\frac{|V|}{k}\right\rceil \leq \left|V_i\right| \leq(1+\varepsilon)\left\lceil\frac{|V|}{k}\right\rceil  .
\end{equation*}

Typically, balanced graph partitioning problems fall under the category of NP-hard problems\cite{feldmann2015balanced}. Solutions to these problems are generally derived using heuristics and approximation algorithms. In our work, we use
% \jwdnote{"use" instead of "introduce", since "introduce" sounds like you are claiming to invent it.} 
a multi-level graph partitioning algorithm\cite{hendrickson1995multi}. The key idea of this algorithm is to reduce the size of the graph by collapsing vertices and edges, partitions the smaller graph, then maps back and refines this partition of the original graph.

(\textbf{Practical Unbalanced Partition - from the community detection perspective}) For the balanced graph partitioning stated above, our primary goal is to evenly distribute nodes to each machine, ensuring that no machine possesses an excessively high workload. Simultaneously, we seek to minimize communication between different machines. However, when each GPU has strong computational capabilities, evenly distributing nodes is no longer our primary goal. 

In this part, our focus is on effectively identifying community structures in the graph. This involves ensuring tight connections in the communities and sparse connections between communities. 
We borrow the idea in community detection \cite{fortunato2010community} and propose an unbalanced partition method based on community detection and spatial information. First, by minimizing modularity \cite{newman2006modularity} through community detection, we obtain $n$ communities with dense internal connections and sparse interconnections. Directly solving the optimal solution in minimizing modularity is hard and time-consuming. Here, we use 
% \jwdnote{ditto last comment} 
the Leiden 
algorithm\cite{traag2019louvain}, which is much faster and yields higher quality solutions. The Leiden algorithm consists of three phases: locally moving nodes, refinement of the partition and aggregation of the network based on the refined partition. For the time complexity, numerical experiments suggest it roughly scales as $O(n)$ or $O(n \log n)$, with $n$ being the number of nodes.
Second, utilizing geographical location information, we calculate the central coordinates for each community. We perform $k$-means clustering to the community centre nodes, then we aggregate $n$ communities to $k$ partitions via the result of clustering. (In our experiment, the community number $n$ is much larger than the GPU number $k$)

(\textbf{Remark: Detailed Explanation of Community Detection}) We introduce the concept of modularity and community detection in the following content as the complement of contents above \cite{lancichinetti2009community}\cite{fortunato2010community}\cite{traag2019louvain}\cite{ahmed2020k}.

Modularity is defined as a value in the range $[-1 / 2,1]$ that measures the density of links inside communities compared to links between communities. For a weighted graph, modularity is defined as:
\begin{equation*}
   Q=\frac{1}{2 m} \sum_{i j}\left[A_{i j}-\frac{k_i k_j}{2 m}\right] \delta\left(c_i, c_j\right) 
\end{equation*}
where
\begin{itemize}
    \item $A_{i j}$ represents the edge weight between nodes $i$ and $j$.
    \item  $k_i$ and $k_j$ are the sum of the weights of the edges attached to nodes $i$ and $j$, respectively.
    \item $m$ is the sum of all of the edge weights in the graph.
    \item $c_i$ and $c_j$ are the communities of the nodes.
    \item $\delta$ is Kronecker delta function ( $\delta(x, y)=1$ if $x=y, 0$ otherwise).
\end{itemize}

Based on calculations, the modularity of a community $c$ can be also represented as:
\begin{equation*}
Q_c=\frac{\Sigma_{i n}}{2 m}-\left(\frac{\Sigma_{t o t}}{2 m}\right)^2    
\end{equation*}
where
\begin{itemize}
   \item $\Sigma_{i n}$ is the sum of edge weights between nodes within the community $c$ (each edge is considered twice)
   \item $\Sigma_{t o t}$ is the sum of all edge weights for nodes within the community (including edges which link to other communities).
\end{itemize}
(\textbf{Remark: Implementation Procedure of Graph Partitioning in multi-GPU computation}) We will describe how to allocate graph information to multiple GPUs using the graph partitioning method from the previous section.
\begin{itemize}
    \item Graph Construction: For a fixed time step $t_k$, we construct the graph by the vehicle path choice from time step $t_{k}$ to $t_{k+1}$. The vertices and the edges of the graph are cities and roads traveled by cars. The weights of vertices and the edges are related to the number of times they are visited by the vehicle.
    \item Graph Partitioning: Implement the graph partitioning methods to the graph constructed before. 
    \item Outlier Detection: For those nodes never visited in the past time period, we allocate them to the nearest subgraph.
\end{itemize}

\begin{figure}[!htbp]
    \centering
    \includegraphics[width=\columnwidth]{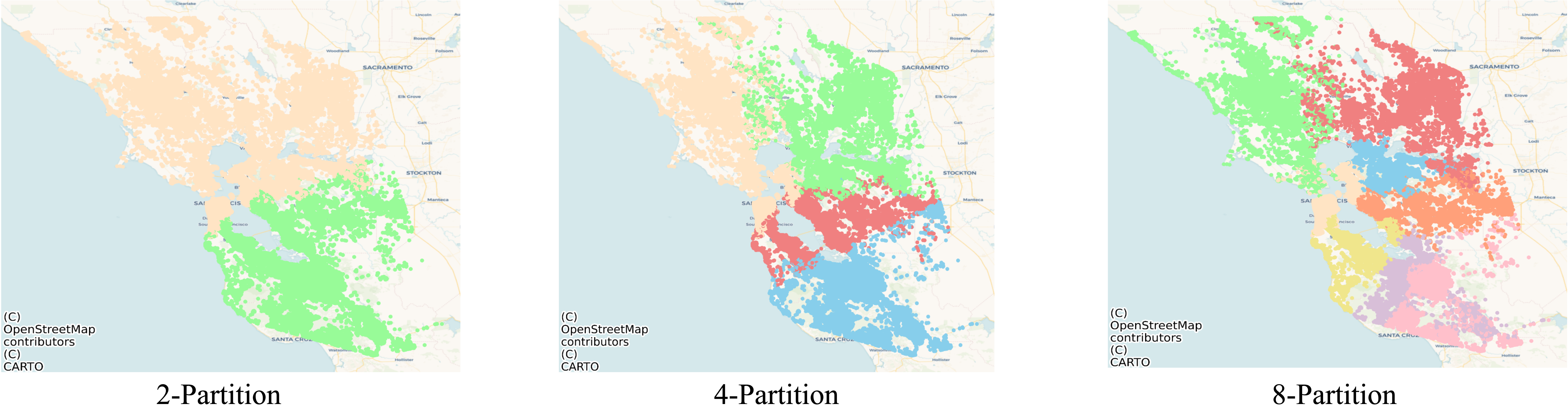}
    \caption{Balanced Graph Partitioning Visualization Based on Half Day Demand}
    \label{fig:Partition_time}
\end{figure}

\begin{figure}[!htbp]
    \centering
    \includegraphics[width=\columnwidth]{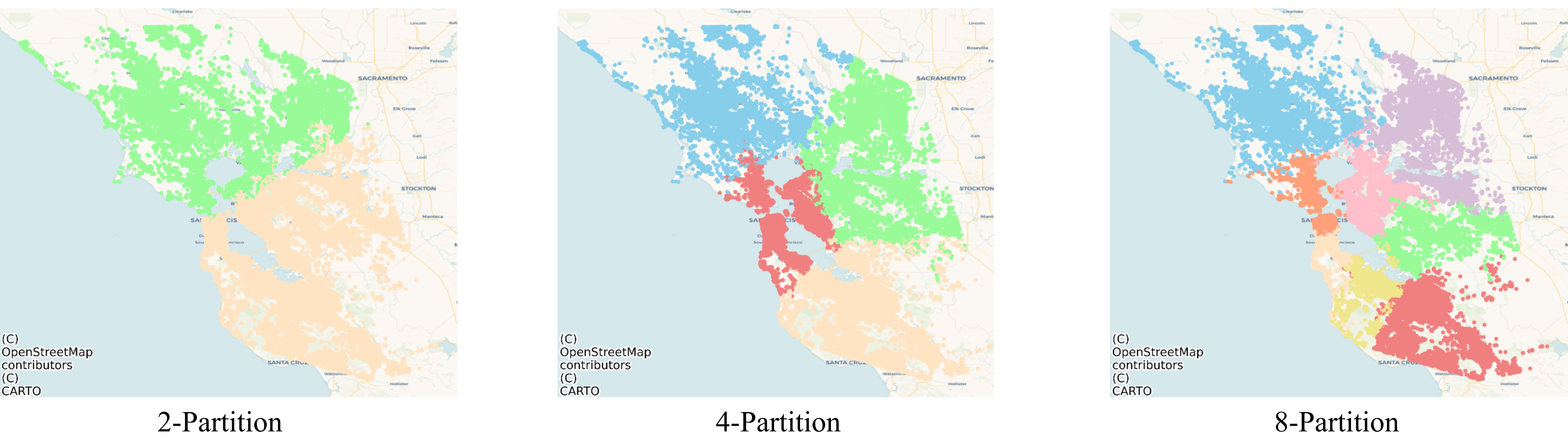}
    \caption{Unbalanced Graph Partitioning Visualization Based on Half Day Demand}
    \label{fig:Partition_time_unbalanced}
\end{figure}

We present an example of partition results for two distinct time periods, showing both balanced and unbalanced graph partitionings in Figures \ref{fig:Partition_time} and \ref{fig:Partition_time_unbalanced}, respectively. These partitions are aligned with the dynamics of the real world traffic flow. Specifically, areas such as the Bay Bridge and Treasure Island are unified within a single cluster due to the incessant flow of traffic in both balanced and unbalanced partition cases because of their heavy traffic throughout the day, underscoring the impracticality of division. In the balanced partition scenario, the map is divided into northern and southern segments by the Bay Bridge area, delineating a clear geographic split. On the other hand, the unbalanced partition strategy, particularly with 4 and 8 clusters, mirrors the Bay Area's administrative boundaries more closely. For the partition featuring four clusters, the blue, green, and orange parts represent North Bay, East Bay, and South Bay, respectively, while the red segment encompasses the San Francisco and Peninsula areas, along with portions of North and East Bay, interconnected by the Golden Gate Bridge and the Bay Bridge. In the scenario with eight clusters, each segment closely approximates the distinct counties within the Bay Area, with the exception that San Francisco and Marin counties are in one segment.
% \jwdnote{It looks like San Francisco and Marin Counties are in one segment, maybe this is what you meant.}

% \begin{figure}[h]
% 	\begin{minipage}{0.32\linewidth}
% 		\vspace{3pt}
% 		\centerline{\includegraphics[width=\textwidth]{image/hc_2clusters.png}}
% 		\centerline{\small 2-Partition}
% 	\end{minipage}
% 	\begin{minipage}{0.32\linewidth}
% 		\vspace{3pt}
% 		\centerline{\includegraphics[width=\textwidth]{image/hc_4clusters.png}}
	 
% 		\centerline{\small 4-Partition}
% 	\end{minipage}
% 	\begin{minipage}{0.32\linewidth}
% 		\vspace{3pt}
% 		\centerline{\includegraphics[width=\textwidth]{image/hc_8clusters.png}}
	 
% 		\centerline{\small 8-Partition}
% 	\end{minipage}
 
% 	\caption{Balanced Graph Partition Visualization Based on Half Day Demand  }
% 	\label{fig:Partition_time}
% \end{figure}

% \begin{figure}[h]
% 	\begin{minipage}{0.32\linewidth}
% 		\vspace{3pt}
% 		\centerline{\includegraphics[width=\textwidth]{image/cd_2clusters.png}}
% 		\centerline{\small 2-Partition}
% 	\end{minipage}
% 	\begin{minipage}{0.32\linewidth}
% 		\vspace{3pt}
% 		\centerline{\includegraphics[width=\textwidth]{image/cd_4clusters.png}}
	 
% 		\centerline{\small 4-Partition}
% 	\end{minipage}
% 	\begin{minipage}{0.32\linewidth}
% 		\vspace{3pt}
% 		\centerline{\includegraphics[width=\textwidth]{image/cd_8clusters.png}}
	 
% 		\centerline{\small 8-Partition}
% 	\end{minipage}
 
% 	\caption{Unbalanced Graph Partition Visualization Based on Half Day Demand  }
% 	\label{fig:Partition_time_unbalanced}
% \end{figure}

% \todo{Something goes wrong with my pdf version. Now checking it...}
 
 % \todo{Formating here @Chonghe}

% \textcolor{red}{Jiaying/Communication}

\subsubsection{Inter-GPU Communication}

\begin{figure}[!htb]
\centering
\includegraphics[scale=0.45]{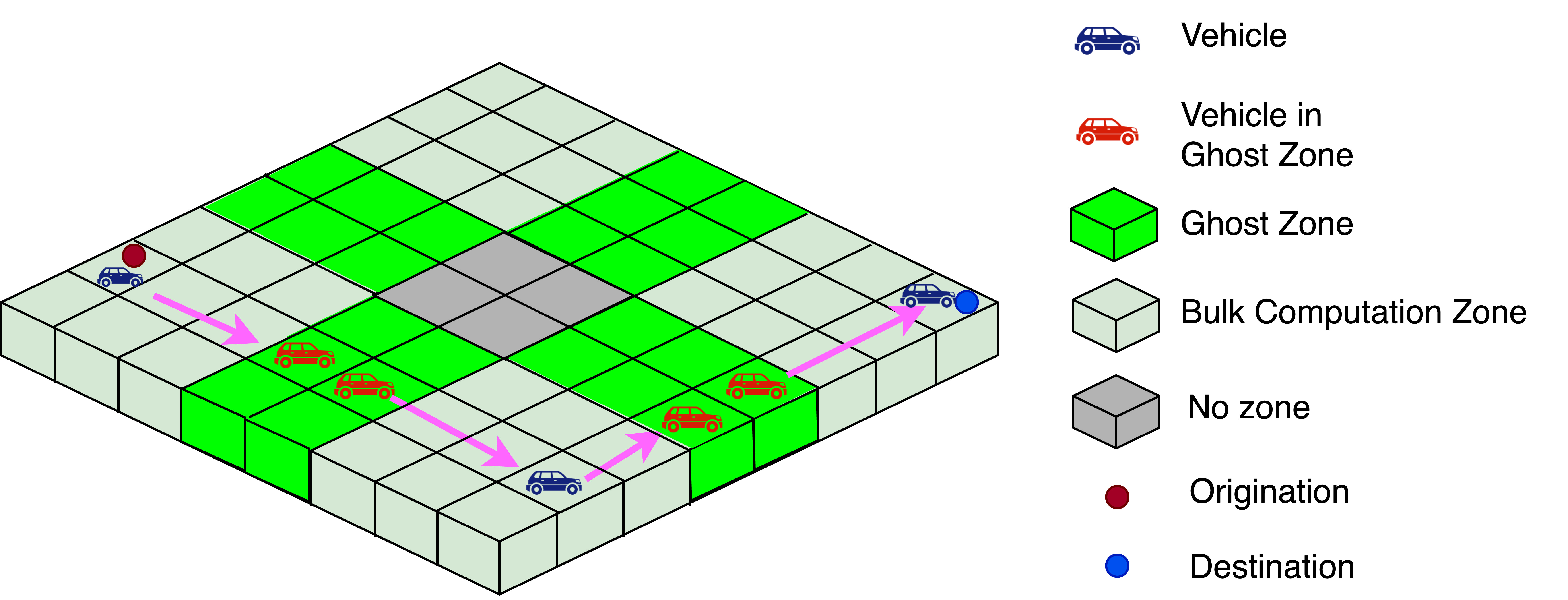}
\caption{Multiple GPU Representation}
\label{multiple-gpu}
\end{figure}

\begin{figure}[!t]
\centerline{\includegraphics[width=\columnwidth]{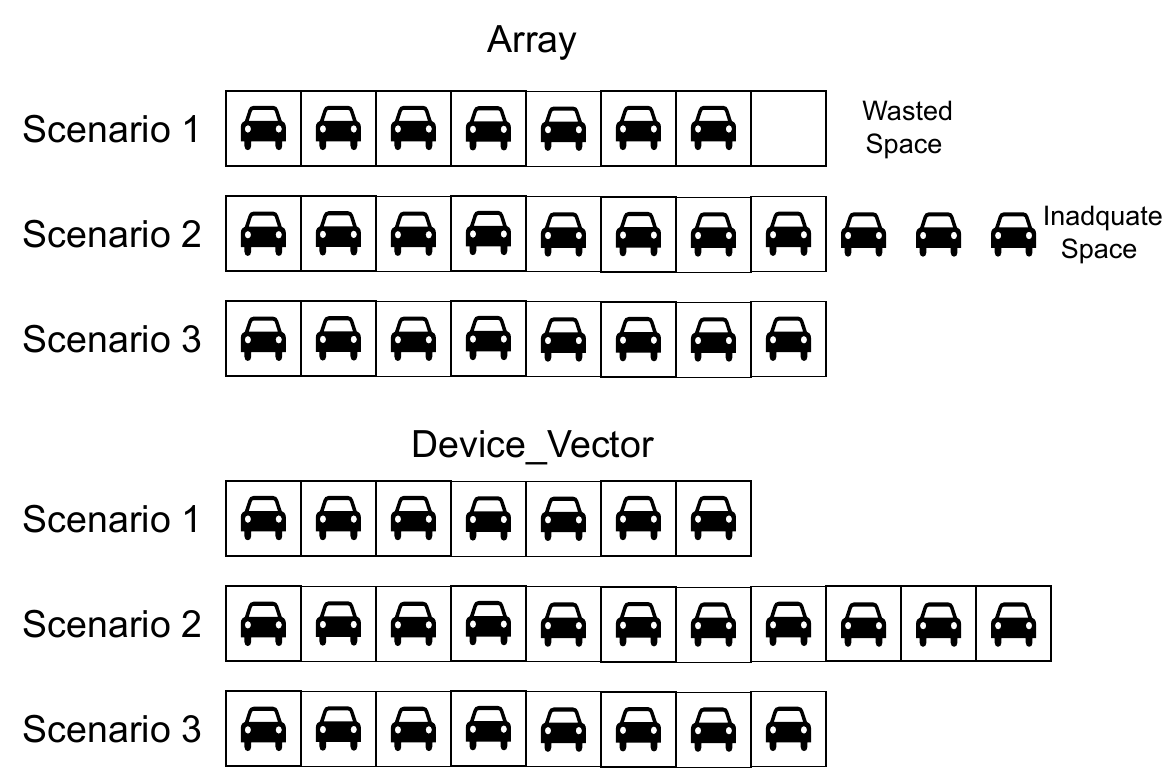}}
\caption{The Advantage of Device\_Vector over Array in Storage of Vehicles}
\label{vehicles_storage_array_vector}
\end{figure}

The additional cost incurred in a multi-GPU \ref{multiple-gpu} setup, compared to a single-GPU system, arises predominantly from the need for communication between GPUs. We facilitate communication between multiple GPUs to handle two types of data: vehicles and road networks within ghost zones. Vehicle data transmission occurs when a vehicle enters a ghost zone, while road networks data communication is necessary to maintain consistency across different GPUs when a vehicle moves within a ghost zone. While the use of arrays to store vehicle data, managed via cudaFree and cudaMalloc, presented an intuitive approach, it revealed significant drawbacks, especially in handling variable data sizes. As illustrated in Scenario 1 of Figure \ref{vehicles_storage_array_vector}, allocating excessive memory leads to inefficiency, whereas insufficient allocation, shown in Scenario 2, poses the risk of array index out of bounds. Moreover, this method was hampered by the time-consuming processes involved in frequent memory allocations and deallocations. To efficiently manage the variable-length vehicle data, which needs to be transferred between GPUs, we employ a more effective approach, utilizing \texttt{device\_vector} from the Thrust library, the CUDA C++ template library. Thrust's \texttt{device\_vector} is designed for GPU contexts and allows dynamic resizing of its contained elements. This is achieved through efficient memory management strategies, which minimize the overhead of reallocating memory when the device\_vector size changes.

Whenever a vehicle is transferred from one GPU to another, we record its original and target data positions. Similarly, when a vehicle needs to be removed from a GPU, we note its data position. A buffer area is designated for tracking vehicles marked for copying or deletion. Each thread responsible for these operations employs atomic operations to identify the buffer position and record the relevant data. This setup facilitates the efficient resizing of the vehicle device\_vector and the management of communication data.

\begin{figure}[!t]
\centerline{\includegraphics[scale=0.80]{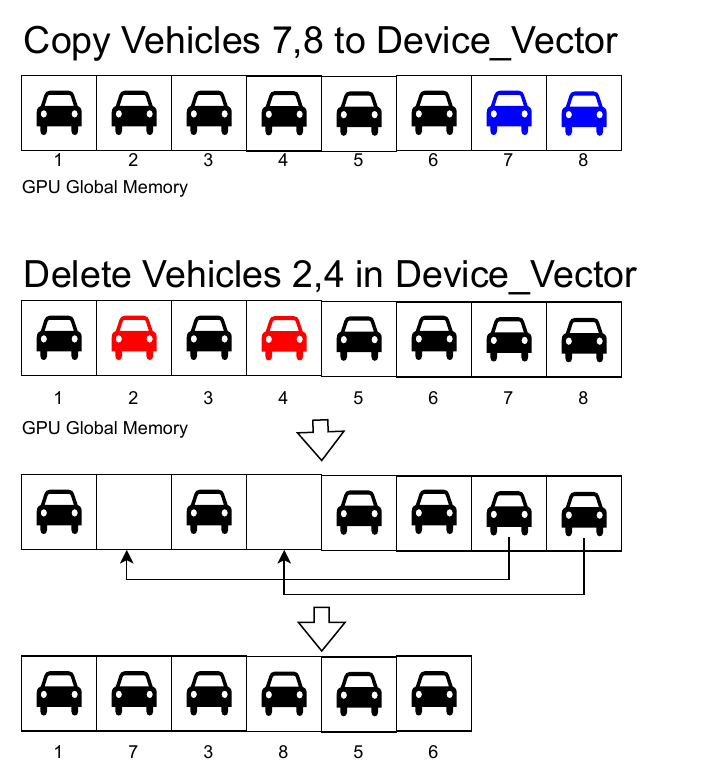}}
\caption{The Demonstration of Vehicle Replication and Deletion}
\label{communication_copy_delete}
\end{figure}

As shown in Figure \ref{communication_copy_delete}, for vehicle replication, threads are launched for each pair of GPUs to handle vehicle data transfer. These threads efficiently employ \texttt{cudaMemcpyPeer} for direct data transfer between GPUs, bypassing the CPU. For vehicle deletions, given the unordered nature of vehicle data storage, the process of data deletion is optimized by moving data from the end of the array 
% \jwdnote{Your use of 3 terms (array, vector and device\_vector) is confusing. Are there just
% two data structures, arrays and device\_vectors,
% and did you mean device\_vector here?} Yes
to the points of deletion. This method is enhanced through parallel execution across multiple threads, significantly speeding up the deletion process. 
% In the case of road network data communication, which is triggered when a vehicle moves within a ghost zone, we maintain consistency across GPUs using a kernel function. This method ensures that all GPUs have up-to-date road network information, critical for accurate traffic simulation.

\begin{figure}[!t]
\centerline{\includegraphics[scale=0.80]{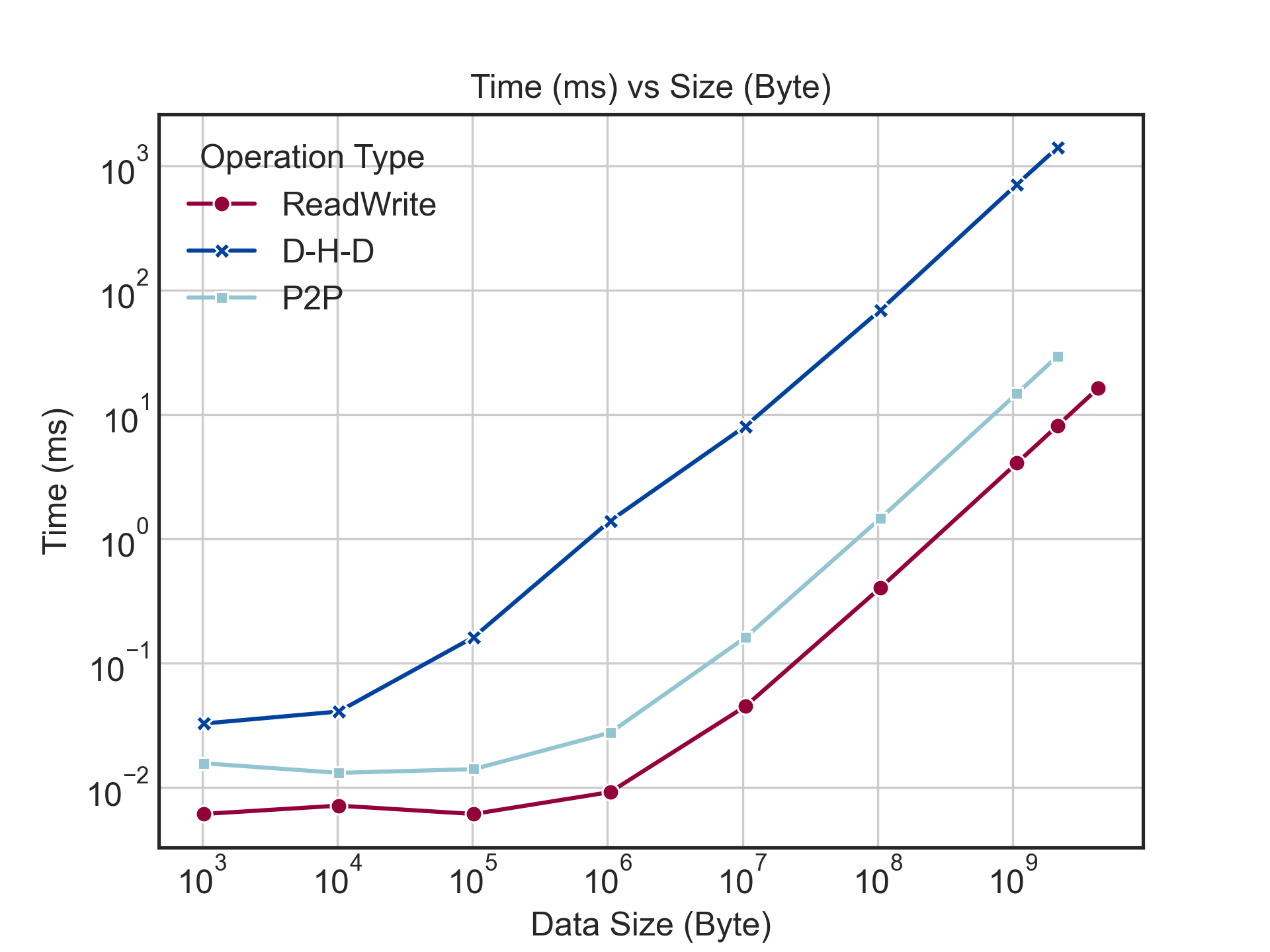}}
\caption{Communication Efficiency}
\label{communication_efficiency}
\end{figure}

Building on the detailed methodologies for vehicle replication and deletion, and road network data consistency, we've established a robust system for managing GPU communication. The following section demonstrates the relatively minimal time impact of communication operations compared to read and write processes from one GPU to its local memory,
% \jwdnote{Do you mean on one GPU to its local memory?}
underscoring the practicality and efficiency of our multi-GPU approach. For read and write operations, the cost was assessed by reading and writing large arrays, where each thread manipulates a single array value. In contrast, communication operations were evaluated by replicating a long array from one GPU to another, using both Device-Host-Device (D-H-D) and Peer-to-Peer (P2P) methods. P2P, especially via NVLink, is claimed faster than D-H-D, though its availability depends on the machine's hardware architecture. Experimental results on V100 GPUs equipped with NVLink revealed that the time cost for P2P communication is approximately threefold that of read-write operations, shown in Figure \ref{communication_efficiency}, indicating a reasonably efficient system. In the upcoming performance section, we will elaborate using the example of San Francisco Bay Area, how graph partitioning techniques and other strategies effectively minimize communication overhead to a mere fraction (about 1\textperthousand) of the total computational load. In summary, while inter-GPU communication does introduce additional time costs, this is relatively minimal compared to the extensive read-and-write operations, thereby justifying the use of multiple GPUs for their significant performance benefits.

\section{Experimental Results}
\label{results}
\subsection{Experimental Data Source}
\label{data}
In this study, we engaged in a collaborative effort utilizing data from the San Francisco County Transportation Authority (SFCTA) \cite{bent2010evaluating}. Based on the San Francisco Chained Activity Modeling Process (SF-CHAMP) \cite{outwater2006san}, a modeling system designed to project transportation patterns across the nine counties of the San Francisco Bay Area. SF-CHAMP 6 leverages observed travel behaviors of San Francisco residents, detailing the socio-economic characteristics and transportation infrastructure of the region to generate key metrics pertinent to transportation and land use planning.

Our analysis focused on a comprehensive dataset encompassing over 23.5 million recorded trips, which is a typical weekday with no significant events or seasonal impacts\cite{chan2023simulating}. Each entry in this dataset was categorized by origin, destination, mode of transportation, and several other critical attributes. A significant portion of our research was devoted to examining car trips, specifically categorized as Single-Occupancy Vehicle (SOV), High Occupancy Vehicle with two occupants (HOV 2), and High Occupancy Vehicle with three or more occupants (HOV 3+). Upon applying these criteria, our scope narrowed down to 17.8 million car trips, constituting 75.7\% of the total dataset. This subset provided a substantial basis for analyzing vehicular movement patterns within the region, offering insights into the predominant vechicle propagation and its implications for urban planning and traffic management.

\subsection{Experimental Results}
\label{Performance}
% % \todo{After checking the correctness, could you run 1GPU, 2GPUs, 4GPUs, 8GPUs, and 16GPUs? @Jiaying}
% \subsection{Instruction roofline model}
% \subsection{Different graph partition technics}
% % \textcolor{blue}{Xuan, Jiaying, Chonghe}
We conducted several numerical experiments to showcase the efficacy of our proposed GPU parallel computing based regional scale traffic simulation framework. Specifically, we first demonstrate the result consistency across different number of GPUs used. Next, we tested the random graph partitioning and the proposed balanced/unbalanced partitions, under peak and non-peak hours, which is used to demonstrate the simulation ability under different scenarios of traffic demands, and give us insights on what kind of graph partitioning to be used under what kind of traffic demand scenarios. Finally, we compare the performance of storing vehicles with Device\_Vector and array.

To ensure a fair comparison between the random partition and the balanced/unbalanced partition methods, we set the observation range from 0-12 hours for both methods. The computation of the vehicle movement would be affected by different ways of graph partitioning whose departure time is within the observation range.

\textbf{1) Performance with different graph partitionings.}
During the benchmark test, we test 3 different graph partitioning methods with 2, 4, and 8 GPUs, which is shown in Table \ref{table:benchmarking}, using gcloud instances with different numbers of V100 GPUs. In order to test the speed of the simulation using different GPUs, we will introduce the notion of ‘Strong Scaling’. Based on Glaser \textit{et al.} \cite{GLASER201597}, in strong scaling, we keep the dataset size (the same demand file in our case) constant but increase the number of processors. 

Table \ref{table:benchmarking} demonstrates a reduction in simulation time as the setup scales from 2 to 8 GPUs in a balanced partition, highlighting an enhancement in multi-GPU simulation performance. However, the low speedup observed when transitioning from 4 to 8 GPUs in the balanced scenario suggests that communication becomes a bottleneck in scenarios where the number of GPUs increases without a corresponding enlargement of the problem size,
% \jwdnote{Why doesn't the low speedup in going
% from 4 to 8 GPUs in the balanced case suggest that communication is
% the bottleneck?}; 
whereas for the unbalanced partition from 4GPUs to 8GPUs, the discrepency of the computations intensity happening in different GPUs becomes a big bottleneck and causes an increase in the simulation time in total.

% \begin{table*}
% \caption{Numerical Results}
%     \begin{tabular}{|c|c|c|c|c|}
% \hline \multicolumn{2}{|c|}{ Simulation Time/ms  } & \multicolumn{1}{|c|}{ 2GPUs } & \multicolumn{1}{|c|}{ 4GPUs } & \multicolumn{1}{|c|}{ 8GPUs} \\
% \hline \multirow{6}{*}{ Algorithm } & Random Partition & Aborted in 80\%	 & Aborted in 80\%  & Aborted in 80\%\\
% \hline & Balanced Partition & 1342463  & 1005954 & 1376888  \\
% \hline & Unbalanced Partition & 1287098 & 1199044 & 1441173	\\
% \hline
% \end{tabular}
% \label{table:benchmarking}
% \end{table*}

\begin{table*}
\caption{Numerical Results}
\centering % This is typically not necessary but ensures that the table is centered

\begin{tabular}{|c|c|c|c|c|}
\hline \multicolumn{2}{|c|}{ Simulation Time/ms  } & 2GPUs & 4GPUs & 8GPUs \\
\hline \multirow{3}{*}{ Algorithm } & Random Partition & Aborted in 80\% & Aborted in 80\% & Aborted in 80\%\\
\cline{2-5} & Balanced Partition & 2498466 & 1483472 & 1269893 \\
\cline{2-5} & Unbalanced Partition & 2014554 & 1679861 & 1783668 \\
\hline
\end{tabular}
% Close the resizebox command here
\label{table:benchmarking}
\end{table*}

\textbf{2) Performance with the adjustment of storing vehicles with Device\_Vector.}
Since we performed an improvement of the data structure from array to Device\_Vector, we also conduct a test to compare the performance of the simulation time, and the result is shown below in Table \ref{table:array&vector}. We use a computer with two NVIDIA A100 40GB GPUs to run the test. The result shows that simulation time for storing vehicles with an array is far slower than storing vehicles with a Device\_Vector. It is mainly because Device\_Vectors usually provide dynamic memory allocation and can be scaled at runtime as needed. Arrays, on the other hand, typically require their size to be determined at compile time. In multi-GPU scenarios, dynamic allocation may allow for more efficient memory usage and better memory allocation policies, reducing memory transfers across GPUs. Using an array also requires large host-to-gpu data transfers rather than Device\_Vector storage.
\begin{table}
\caption{Comparison of storing vehicles with Device\_Vector over array}
\centering

\begin{tabular}{|c|c|c|c|c|}
\hline
\multicolumn{2}{|c|}{} & \multicolumn{3}{c|}{SF bay area 9 counties traffic} \\ 
\hline
\multicolumn{2}{|c|}{Graph Partitioning method} & Random & Balanced & Unbalanced\\ 
\hline
\multirow{2}{*}{Way of storing} & Device\_Vector & Aborted & 1342463 & 1287098 \\
\cline{2-5}
& Array & Aborted & 70725162 & 71131945 \\
\hline
\end{tabular}
\label{table:array&vector}
\end{table}

% \textbf{4) GPU utilization efficiency with simulation time goes on}
% @jiaying take a look 

\begin{figure}[H]
\centerline{\includegraphics[width=1.0\columnwidth]{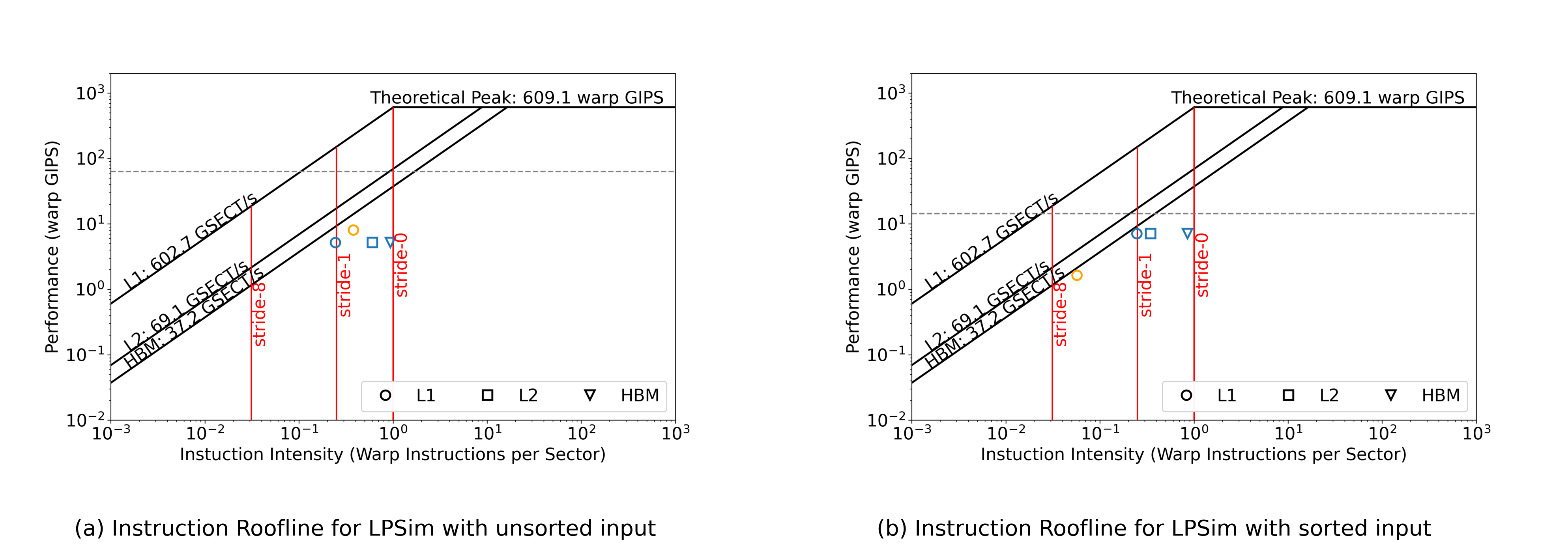}}
\caption{
The Instruction Roofline for our simulator. The \textbf{\textcolor{black}{black line}} represents the theoretical maximum instruction throughput on a warp basis, calculated based on the A100’s architecture, which comprises 108 Streaming Multiprocessors (SMs) with four sub-partitions each, integrating a single warp scheduler capable of dispatching one instruction per cycle within each sub-partition. Consequently, the theoretical maximum instruction throughput on a warp basis is calculated as 108×4×1×1.41 GHz, amounting to 609.12 GIPS. Performance metrics are determined by the methodology of Yang et al. (2019), normalized to warp-level, and rescaled into gigasectors per second (GSECT/s) based on the sector size. The \textbf{\textcolor{red}{red line}} indicates the efficiency of memory access patterns. A warp-level load can result in 1 to 32 sector transactions, making the x-axis digit meaningful. A value of 1 indicates “stride-0” access, where all threads in a warp reference a single memory location, generating only one transaction. Conversely, scenarios like random access or striding beyond 32 bytes (“stride-8”) can lead to the maximum 32 transactions. For unit-stride (“stride-1”) access, typical of FP32 or INT32 operations, the global load/store (LD/ST) intensity stands at 1/4. \textbf{\textcolor{blue}{Blue dots}} represent non-predicated instructions for each level of the memory hierarchy, highlighting the limited data locality and speedup from sorting. \textbf{\textcolor{orange}{Gold dots}} are non-predicated loads per global memory access, highlighting the loss in near-unit stride access from sorting. The dotted line represents the total (including predicated) instruction throughput. Proximity of blue dots to the highlighted line quantifies the impact of predication.
}
\label{roofline_result}
\end{figure}

\textbf{3) Roofline Model Result.}
% @jiaying
% 1. mention the machine we used
% 2. put the results of roofline model and interpret it
Figure \ref{roofline_result} shows the Instruction Roofline applied to NVIDIA’s A100 for our simulator analysis. Blue dots represent non-predicated instructions for each level of the memory hierarchy, highlight the limited data locality, and speedup from sorting. Gold dots are non-predicated loads per global memory access and highlight the loss in near-unit stride access from sorting.  The dotted line represents the total (including predicated) instruction throughput.  Proximity of blue dots to the highlighted line quantifies the impact of predication. The A100's architecture, comprising 108 Streaming Multiprocessors (SMs) with four sub-partitions each, integrates a single warp scheduler capable of dispatching one instruction per cycle within each sub-partition. Consequently, the theoretical maximum instruction throughput on a warp basis is calculated as 108×4×1×1.41 GHz, amounting to 609.12 GIPS. We leverage Yang \textit{et al.}'s methodology \cite{Yang2019HierarchicalRA} for measuring GPU memory bandwidths but rescale into gigasectors per second (GSECT/s) based on the sector size. We record the number of instructions executed on the thread level, normalized to warp-level by dividing by 32. This instruction count is then divided by the corresponding sector count in data movement to determine the instruction intensity. Performance metrics are calculated by dividing this instruction count by the kernel execution time. The resulting blue markers determined by these calculations fall beneath the sloping roofline, illustrating that the program is memory-bound with room to boost instruction intensity for better performance. Moreover, the disparity between the L1 marker and the L2 and HBM markers underscores efficient L1 data reuse, but not the L2.

% Thread predication in GPUs deactivates threads not following a branch, impacting performance by limiting active thread count in a warp. Since a warp processes a single instruction collectively, the fraction of active threads directly influences efficiency. 

% The dotted line in figures represents the maximum warp-level performance, with actual performance falling below this threshold. As is shown in figure \ref{roofline_result}(a), the dots are well below the dotted line indicating a 10× loss in performance due to thread predication. To address such a significant performance loss, we sorted the input OD pairs by departure time. This approach is depicted in the new roofline results shown in Figure \ref{roofline_result}(b), where the gap between the performance markers and the theoretical maximum (dotted line) noticeably narrows, reducing the performance loss from a factor of 10 to just 2. This improvement is largely due to the sorting making threads within the same warp more likely to execute similar instructions, significantly reducing the inefficiencies associated with thread predication. For simulations of vehicles within the same area, minimizing the number of instructions is preferable as it leads to reduced resource wastage, thereby accelerating the computation process.

Before we dive into \textbf{Performance Analysis with Roofline Model}, we will first introduce the concept of thread predication. Thread predication in GPUs is a mechanism designed to handle conditional branches, allowing the SIMD architecture to continue operating even when threads within a warp follow different execution paths. This section provides a detailed explanation of how thread predication works and its impact on performance.

\textbf{Concept of Thread Predication: }In GPU architecture, a warp is a group of threads that execute instructions simultaneously. When a conditional branch (e.g., an \texttt{if-else} statement) occurs, different threads in the same warp might need to take different execution paths based on their individual conditions. Since a warp executes the same instruction at the same time across all its threads, this presents a challenge.

\textbf{Mechanism of Thread Predication: }When the warp encounters a conditional branch, it must execute both paths of the branch to ensure all threads complete their respective instructions. The GPU uses predication to manage this. It deactivates (predicates off) the threads that do not meet the condition for the current branch, while the active threads execute the instructions for that branch. Once this is done, the GPU switches to the other branch, activating the previously deactivated threads and deactivating those that have already executed their instructions. Consequently, the warp processes each branch sequentially, with only a subset of threads active at any one time.

For instance, consider the following example:

\begin{verbatim}
if (condition) {
    // Branch A: executed by threads where 'condition' is true
    doSomething();
} else {
    // Branch B: executed by threads where 'condition' is false
    doSomethingElse();
}
\end{verbatim}

If the \texttt{condition} is true for half of the threads in a warp, those threads will execute \texttt{doSomething()} while the other half are deactivated. Then, the roles are reversed, and the second half executes \texttt{doSomethingElse()}.

\textbf{Impact on Performance: }The impact of thread predication on performance is significant. Since only a portion of the warp's threads are active at any given time, the GPU's parallel processing capability is underutilized. For instance, if only half the threads are active, the warp operates at 50\% efficiency. Each branch must be executed separately for the different sets of active threads, effectively doubling the instruction count for divergent branches. This sequential processing of branches increases the overall execution time, leading to a performance bottleneck when there is significant divergence.

\textbf{Performance Analysis with Roofline Model: }Thread predication deactivates threads not following a branch, impacting performance by limiting the active thread count in a warp. Since a warp processes a single instruction collectively, the fraction of active threads directly influences efficiency. 

Figure \ref{roofline_result}(a) shows the performance analysis using the Roofline model, where the dots are well below the dotted line representing the maximum warp-level performance. This indicates a 10× loss in performance due to thread predication. To address this significant performance loss, we sorted the input OD pairs by departure time. The new roofline results, shown in Figure \ref{roofline_result}(b), reveal that the gap between the performance markers and the theoretical maximum (dotted line) noticeably narrows, reducing the performance loss from a factor of 10 to just 2. This improvement is largely due to the sorting making threads within the same warp more likely to execute similar instructions, significantly reducing the inefficiencies associated with thread predication. Minimizing the number of instructions for simulations of vehicles within the same area is preferable as it leads to reduced resource wastage, thereby accelerating the computation process.

\begin{table}[H]
\caption{Memory Access and Execution Time Before/After Sorting
%\swwnote{do you have an efficiency metric like instructions per car -or- cars per kilo-instructions?}
}
\centering
\scalebox{0.92}{
\begin{tabular}{|c|c|c|c|}
\hline
Version & Load/Store Instruction Count & Sectors Count & Time (s) \\ \hline
Unsorted & 5,533,942 & 14,665,154 & 195.28 \\ \hline
Sorted & 830,574 & 14,547,567 & 170.75 \\ \hline
\end{tabular}
}
\label{memory_access_sort_info}
\end{table}

When a warp accesses global memory, the thread access pattern within that warp is critical, as inefficient memory access patterns can lead to additional, unnecessary transactions, reducing performance. A warp-level load can result in 1 to 32 sector transactions, making the x-axis digit meaningful. A value of 1 indicates "stride-0" access, where all threads in a warp reference a single memory location, generating only one transaction. On the opposite end, scenarios like random access or striding beyond 32 bytes ("stride-8") can lead to the maximum 32 transactions, showcasing the spectrum of memory access efficiency. For unit-stride ("stride-1") access, typical of FP32 or INT32 operations, the global load/store (LD/ST) intensity stands at 1/4. Our model lies between stride-1 and stride-0 which is indicating a good global memory access pattern of our model.  However, as illustrated with the gold dots
% \swwnote{with the gold dots} 
in Figure \ref{roofline_result}(b), the sorted version reveals a deterioration in the memory access pattern. This observation is supported by the statistics of instruction count and sector numbers presented in Table \ref{memory_access_sort_info}. After sorting, the total number of instructions decreased, yet the number of sectors accessed  decreased only slightly. This discrepancy suggests that the compiler optimizations may have optimized away many instructions in the sorted program. It is important to emphasize that the Roofline model, while useful, should be considered a supplementary tool that benefits from being paired with runtime and scalability analysis. In our case, despite the memory access pattern requiring further improvement, the reduction in execution time post-sorting confirms the beneficial impact of sorting.

\textbf{4) Simulation time with different demand sizes.}
We selected demand sizes of 3M, 6M, 12M, 20M, and 24M demands randomly from the full-day SFCTA demand as described in Section \ref{data} to be tested on AWS P3 instances with various numbers of NVIDIA V100 GPUs. We can see that for all the demand pairs, we gained time benefits from increasing GPU numbers from 1 to 2. And for demand of 12M and 20M, we can see time benefit from using 1 GPU all the way to 4 GPUs. And for 24M demand, it can only be simulated with more than 4 GPUs.

Table \ref{table:benchmarking} demonstrates a reduction in simulation time as the setup scales from 1 to 4 GPUs in both balanced and the unbalanced partition with demands of more than 12M, highlighting an enhancement in multi-GPU simulation performance. However, the increased runtime observed when transitioning from 4 to 8 GPUs in the balanced and unbalanced scenario suggests that communication becomes a bottleneck in scenarios where the number of GPUs increases without a corresponding enlargement of the problem size.
% \jwdnote{Why doesn't the low speedup in going
% from 4 to 8 GPUs in the balanced case suggest that communication is
% the bottleneck?}; 

% \begin{table*}
% \caption{Numerical Results}
%     \begin{tabular}{|c|c|c|c|c|}
% \hline \multicolumn{2}{|c|}{ Simulation Time/ms  } & \multicolumn{1}{|c|}{ 2GPUs } & \multicolumn{1}{|c|}{ 4GPUs } & \multicolumn{1}{|c|}{ 8GPUs} \\
% \hline \multirow{6}{*}{ Algorithm } & Random Partition & Aborted in 80\%	 & Aborted in 80\%  & Aborted in 80\%\\
% \hline & Balanced Partition & 1342463  & 1005954 & 1376888  \\
% \hline & Unbalanced Partition & 1287098 & 1199044 & 1441173	\\
% \hline
% \end{tabular}
% \label{table:benchmarking}
% \end{table*}

% \begin{table*}
% \caption{Numerical Results}
% \centering % This is typically not necessary but ensures that the table is centered
% \begin{tabular}{|c|c|c|c|c|}
% \hline \multicolumn{2}{|c|}{ Simulation Time/ms  } & 2GPUs & 4GPUs & 8GPUs \\
% \hline \multirow{3}{*}{ Algorithm } & Random Partition & Aborted in 80\% & Aborted in 80\% & Aborted in 80\%\\
% \cline{2-5} & Balanced Partition & 2498466 & 1483472 & 1269893 \\
% \cline{2-5} & Unbalanced Partition & 2014554 & 1679861 & 1783668 \\
% \hline
% \end{tabular}
% % Close the resizebox command here
% \label{table:benchmarking}
% \end{table*}

\begin{table*}
\caption{Simulation Times for Different Partitions}
\centering
\begin{tabular}{@{}lcccccc@{}}
\toprule
Demand Size & Partition Type & \multicolumn{4}{c}{Simulation Time with Different Number of GPUs} \\
\cmidrule(r){3-6}
& & 1 GPU & 2 GPUs & 4 GPUs & 8 GPUs \\
\midrule
\multirow{2}{*}{3,000,000} 
    & Balanced Partition & 381997 & 353399 & 439829 & 689898 \\
    & Unbalanced Partition & 381997 & 343435 & 446019 & 678993 \\
\addlinespace
\multirow{2}{*}{6,000,000} 
    & Balanced Partition & 752695 & 610608 & 653128 & 928397 \\
    & Unbalanced Partition & 752695 & 597259 & 632208 & 868127 \\
\addlinespace
\multirow{2}{*}{12,000,000} 
    & Balanced Partition & 1535082 & 1182821 & 1087738 & 1306006 \\
    & Unbalanced Partition & 1535082 & 1134121 & 1111049 & 1304821 \\
\addlinespace
\multirow{2}{*}{20,000,000} 
    & Balanced Partition & 2554437 & 1746744 & 1459479 & 1639141 \\
    & Unbalanced Partition & 2554437 & 1768196 & 1717089 & 1635972 \\
\addlinespace
\multirow{2}{*}{24,000,000} 
    & Balanced Partition & n/a & n/a & 1744777 & 1938588 \\
    & Unbalanced Partition & n/a & n/a & 2011620 & 1928708 \\
\bottomrule
\end{tabular}
\label{table:benchmarking}
\end{table*}

\section{Conclusion}
\label{Conclusion}
This paper introduces LPSim, a cutting-edge traffic simulation framework that leverages multi-GPU computation for enhanced performance and efficiency. Unlike traditional traffic simulation models, LPSim integrates graph partitioning methods tailored for multi-GPU environments, ensuring a near-optimal resource utilization and faster processing times.

A key innovation of LPSim is its multi-GPU computation strategy, which allocates graph information across multiple GPUs and manages the spatio-temporal data efficiently. This approach accelerates the computation of the system, allowing for the simulation of more complex and larger traffic networks than the previous work. The communication component of LPSim, especially in multi-GPU setups, is carefully designed to handle vehicle movements across different GPUs, ensuring data consistency and simulation reliability.

The experimental results, validated against real-world data, demonstrate LPSim's accuracy \cite{jiang2024drbo} in replicating traffic dynamics. The framework's performance analysis reveals significant advantages in using multiple GPUs over a single GPU setup, including scalability and efficiency in handling large-scale traffic simulations.

Future enhancements of LPSim are threefold. 

Firstly, from the multimodal perspective, we will focus on extending its capabilities to multimodal traffic scenarios, further bridging the gap between theoretical traffic models and practical traffic management applications. This will deepen our understanding of complex routing challenges within simulations, paving the way for more comprehensive and accurate traffic modeling. Although we have not tested the multimodal systems on LPSim, we have gotten the bike network from Open Street Map \cite{map2014open}, and the SFCTA data \cite{bent2010evaluating} we have provides the bike mode.
 We plan to use these datasets to test multimodal systems in the future. This progressive approach will help in the field of traffic simulation and smarter management, contributing to smarter, more responsive urban planning strategies. 

Secondly, from the theoretical design and analysis perspective, we are passionate about capturing the dynamics of the simulation system and designing a more efficient graph partitioning strategy. And we'll explore the possibility of using shared memory for better data locality to improve modal. The computation model given in this paper will be further explored. In addition to refining the model, more detailed calibration and routing algorithms are crucial for analyzing real-world results. Our team is actively investigating the application of dynamic traffic rerouting strategies. To ensure the accuracy and validity of our model, we will leverage Uber movement data \cite{sun2020uber} for calibration and validation purposes. 

Thirdly, acknowledging the inefficiencies of storing vehicle and road network information in global memory, we consider partitioning each GPU into multiple segments, wherein related data would be stored in shared memory. This approach is anticipated to facilitate faster data retrieval and processing, as shared memory access times are comparable to those of the L1 cache, thereby potentially reducing the latency associated with global memory access. However, this strategy introduces the potential for additional computational and communication overhead. We plan to rigorously evaluate the trade-offs between improved data access speeds and the increased complexity of managing shared memory spaces.

In the end, the framework's user-friendly design, with a one-click Docker setup, makes it accessible to a broad range of users, from researchers to urban planners.
The open-source nature of LPSim not only highlights its potential for broad adoption but also invites ongoing enhancements and innovations from the global community. Looking ahead, there are plans to integrate LPSim into a variety of practical applications, notably in urban traffic management and the development of smart cities. This integration is poised to significantly enhance the efficiency and sustainability of urban infrastructures.

\bibliography{mybibfile}

\end{document}